\documentclass{svproc}
\usepackage{graphicx}
\usepackage{latexsym}
\usepackage{color}
\usepackage{makeidx}
\begin{document}

\authorrunning{T. Gubiec. J. Klamut, and R. Kutner}
\titlerunning{Multi-phase long-term autocorrelated diffusion}
\title{Multi-phase long-term autocorrelated diffusion: \\
Stationary continuous-time Weierstrass \\
walk vs. flight}
\author{Tomasz Gubiec, Jaros{\l}aw Klamut, \and Ryszard Kutner}
\institute{Faculty of Physics, University of Warsaw, Pasteur Str. 5, PL-02093 Warsaw, Poland}

\date{Received: date / Revised version: date}

\maketitle
\abstract{
In this paper we are examining diffusion properties of stationary 
continuous-time Weierstrass walk (CTWW). We are showing it is 
a multi-phase representation of the L\'evy walk. The hierarchical spatial-temporal coupling, combined with coupling between dynamic variables define the CTWW process. The walker moves here with a piecewise constant velocity between trajectory turning points. We have found the diffusion phase diagram of the CTWW consisting not only of anomalous non-Gaussian or non-fBm phases but also Brownian yet 
non-Gaussian ones. We compare the diffusion phase diagram of the stationary CTWW with the corresponding hierarchical continuous-time Weierstress flight (CTWF). The instantaneous jumps between trajectory turning points preceded by waiting define the CTWF process. It is a hierarchical representation of the L\'evy flight. We have found the diffusion phase diagram of the CTWF to be a small part of the corresponding CTWW one.
}

\section{Historical sketch and inspirations}\label{section:hypothesis}

\subsection{Initial remarks}

By the pioneering work published in the year 1965 \cite{MonWeis}, physicists Eliott W. Montroll and George H. Weiss introduced the concept of continuous-time random walk (CTRW) as a way to achieve the inter-event time continuous and fluctuating\footnote{Let us incidentally comment that the term `walk' in `continuous-time random walk' is commonly used in the generic sense comprising two concepts namely, both the walk (associated with the finite velocity of the process) and flight (associated with an instantaneous displacement of the process). Thus, it has to be specified in concrete consideration what kind of a process we have.}. This stochastic process is characterized by the waiting-time and displacement joint distribution\footnote{This distribution can be also called pausing- or inter-event time and displacement distribution. Previously, the incomplete term `waiting-time distribution' (WTD) was used.} (WTDD), giving an insight into the process' details in various scales. The WTDD allows the description of both Debye (exponential) and non-Debye (slowly decaying) relaxations as well as regular and anomalous transport and diffusion \cite{KlafZum,ZubDenKlaf}. It is promising that the WTDD is also used to characterize evolving complex systems \cite{ParKanScar}.
 
The publication \cite{MonWeis} mentioned above stirred up a little attention of a community, until in the early seventies Harvey Scher and collaborators applied the CTRW technique to the anomalous transport and diffusion in amorphous materials \cite{ScRev}. The article of Shlesinger \cite{MFSchle} presents a thorough review of the history and antecedents of the CTRW. 

In subsequent years, the importance of the formalism soared with an increasing number of works generalizing and applying the CTRW with the number of cites climbing to several tens of thousands. The CTRW constitutes an extremely powerful, albeit relatively simple formalism to approach and eventually solve a countless number of problems in many areas of physical, natural, and even socio-economic sciences. In this regard, see our review article \cite{KutMasol}.

Presumably, the most commonly used CTRW formalism was developed by Scher and Lax in terms of recursion relations \cite{ScLax0,SchLax,MonWes,WeiRub,MonShle,Weis}. In this context, the distinction between separable and non-separable WTDDs, which is key for present work, was introduced \cite{HauKehr0}. A thorough analysis of the non-separable one \cite{MonMasol,KKK,Kut0} took into account dependencies between many correlated consecutive particle displacements and waiting (or inter-event) times.

The canonical version of the CTRW formalism concerning transitions between different sites and states by using the recursion relations is equivalent to the generalized master equation (GME). This is since one-to-one transformation between WTDD and memory kernel was clearly established \cite{KenMonShle1,KenKno,Kenk,LandMonShle,LandShle1,Gill,CacWio,MasWeis,KlafSil}. This creates the foundation of our work because memory can lead to the anomalous and non-Gaussian diffusion.
 
Although initially, CTRW was a kind of renewal theory, Tunaley was able to modify it by preparing the class of initial (averaging) WTDD, crucial for this work. 
Such a modification makes CTRW time-homogeneous \cite{Tun,Tun3} and enables to consider CTRW a semi-Markov process \cite{MeStra,BarLim}. 
Thus the application of the fundamental Wiener-Khinchin theorem (relating autocorrelation function to power spectra) became possible. 

In principle, the CTRW is fundamentally different from the regular random flight or walk models as their probability densities in the long-time limit scales in a non-Gaussian way, being a severe and inspirational extension of the Gaussian one. 
Thus, the CTRW became a foundation of anomalous (dispersive, non-Gaussian) transport and diffusion \cite{Weis,ScheMon,PhisScher} opening the modern and trendy segment of statistical physics, as well as condensed and soft matter physics, stimulating their very rapid expansion even outside of the traditional statistical physics (including statistical physics of open systems \cite{EsLin}) \cite{BouGeo,SokKlaf}. 
In this work, we consider multi-phase~diffusion by analyzing~diffusion phase diagrams.

\subsection{Elements of L\'evy random motion} 

Thanks to its versatility, the CTRW found numerous vital applications in many fields ranging from biology through telecommunication to finance, including econometrics and economics, and even to speech recognition. 
The CTRW found countless applications in many other areas that are still developing, such as aging of glasses, a nearly constant dielectric loss response in structurally disordered ionic conductors as well as in modeling of hydrological models and earthquakes. 

When a particle performs a CTRW of the flight (jump or hop) type then during its evolution, it makes instantaneous jumps alternating with waiting-events (or rests). The CTRW formalism allows to combine both of these particle states, offering an abundant diffusion phase diagram or several scaling regimes. Moreover, the CTRW formalism can be extended to assume walks with finite velocity instead of instantaneous jumps. Such a kind of a model we call the L\'{e}vy walk interrupted by rests \cite{ZubDenKlaf,KlaftZum1,ZabChuk,KlafSok}. The presence of finite particle velocity there significantly increases the flexibility of this kind of models. However, at the same time it makes it more difficult to find their analytical solutions if those exist. The standard versions of L\'{e}vy walk model, i.e., without rests, were also intensively developed assuming fixed particle velocity \cite{KlafBluShl,ZumKlaf} or varying velocity, e.g., according to the self-similar hierarchical structure \cite{KutSwit}. The present work is an extension of them \cite{KutSwit}.

There are several generalizations of the L\'{e}vy walk model \cite{ZubDenKlaf} which assume that particle velocity can vary randomly \cite{BarKlaf,ZabSchSta} or by some other rules \cite{ZabSchSta}. Among those, L\'{e}vy walk model with random velocity (LWRV) and the one with weakly fluctuating velocity caused by the active environment \cite{DenZabHang,ZabDenHang,ZabDenHang1} are inspiring. The LWRV model was applied, for instance: (i) in physical problems of two-dimensional turbulence \cite{TonGol,MML,Chuk}, (ii) as a model of velocity distribution in kinetic theory \cite{BMM,TBE}, (iii) in plasma physics \cite{HMD,MTS}, and (iv) in some non-extensive statistics \cite{Tsal,Taketal}. 

The finite velocity of walking particles constitutes random walk models more general than the erratic flight (jump) ones and bring them closer to the physical principles. Indeed, more fluctuations are responsible for real-life aspects, which the CTWW formalism takes into account in contrast to the CTWF one. These two formalisms are systematic compared with each other in this work giving a contribution to a stream of works on the subject \cite{KutPekSznaj0,EliaKlaft,KlafSok,SaCheMe}.

\section{Stationary continuous-time Weierstrass walk vs. flight: Definitions}\label{section:defs}

\begin{figure}
\begin{center}
\bigskip
\includegraphics[width=120mm,angle=0,clip]{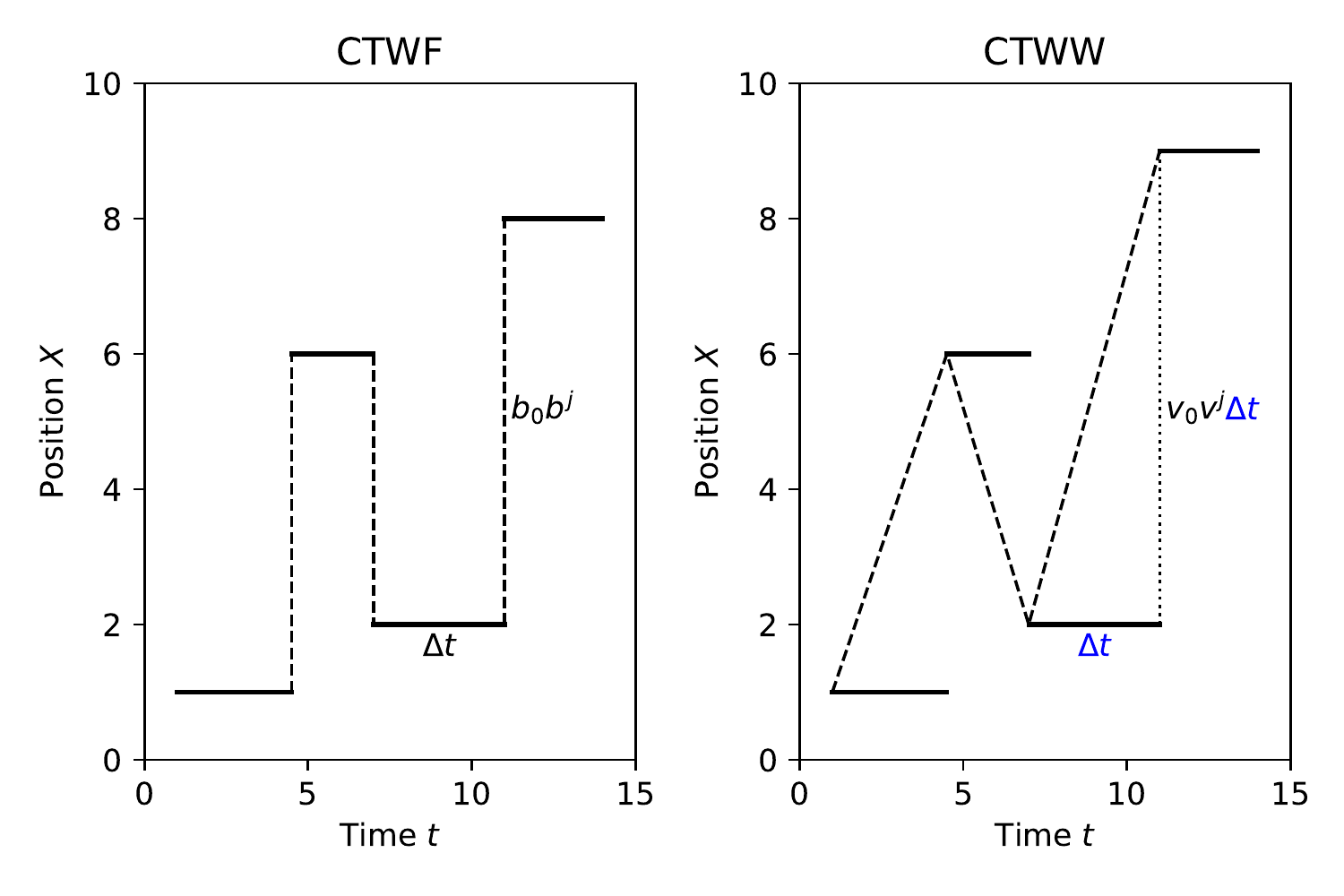}
\caption{Comparison of typical one-dimensional trajectories of CTWF (left plot) and CTWW (right plot) processes. Horizontal fluctuating time intervals $\Delta t$ (substantial horizontal segments) denote process time-lags between successive turning points. In the case of the CTWF process, it means waiting- or pausing-time, whereas in the case of the CTWW process, it means the inter-event time needed to overcome segments $\pm v_0v^j\Delta t$ of the trajectory between successive turning points. In the case of CTWF, parts $\pm b_0b^j$ of the path are traveled abruptly (dashed vertical segments of straight lines). Whereas, in case of the CTWW, the dashed sloped fragments of straight lines mimic the walk at a constant velocity value of $v_0v^j$. Recall that both $j$ and $\Delta t$ are random variables (discussed in Sec. \ref{section:defs}), which can be different for each pair of successive turning points.}
\label{figure:flucWF}
\end{center}
\end{figure}

When we consider the continuous-time Weierstrass walk we allow the walker to move with a constant velocity between the successive trajectory turning points \cite{KutSwit,SChSPM,Kutner1}. That is, the walker (process) velocity is a piecewise constant function of time.
Fig. \ref{figure:flucWF} shows a comparison of typical CTWF and CTWW trajectories' segments. 

Both processes we define by Eqs. (\ref{rown:psixDtF}) and (\ref{rown:psixDtW}). The location of turning points of the CTWF is defined utilizing instantaneous single-step displacements $x=\pm b_0b^j=\pm v_0v^j\langle \Delta t_j\rangle $, where $\langle \Delta t_j\rangle =\tau _0\tau ^j, j=0,1, \ldots , b_0=v_0\tau _0, b=v\tau $. In case of the CTWW, the corresponding turning points are defined by using temporary single-step displacements $x=\pm v_0v^j\Delta t_j$, while the stochastic variable $\Delta t_j=\Delta t$, where $\Delta t$ is in each step drawn from the stationary exponential distribution characterized by relaxation time $\tau _0\tau ^j$.
For both processes, the index $j$ of the hierarchy's level is drawn from the geometric distribution, 
\begin{eqnarray}
w(j)= \left(1-\frac{1}{N}\right)\frac{1}{N^j}. 
\label{rown:wj}
\end{eqnarray} 
We obtained this distribution simply assuming that the ratio $\frac{w(j+1)}{w(j)}$
is stationary, i.e., assuming its independence from index $j$.

\begin{figure}
\begin{center}
\bigskip
\includegraphics[width=119mm,angle=0,clip]{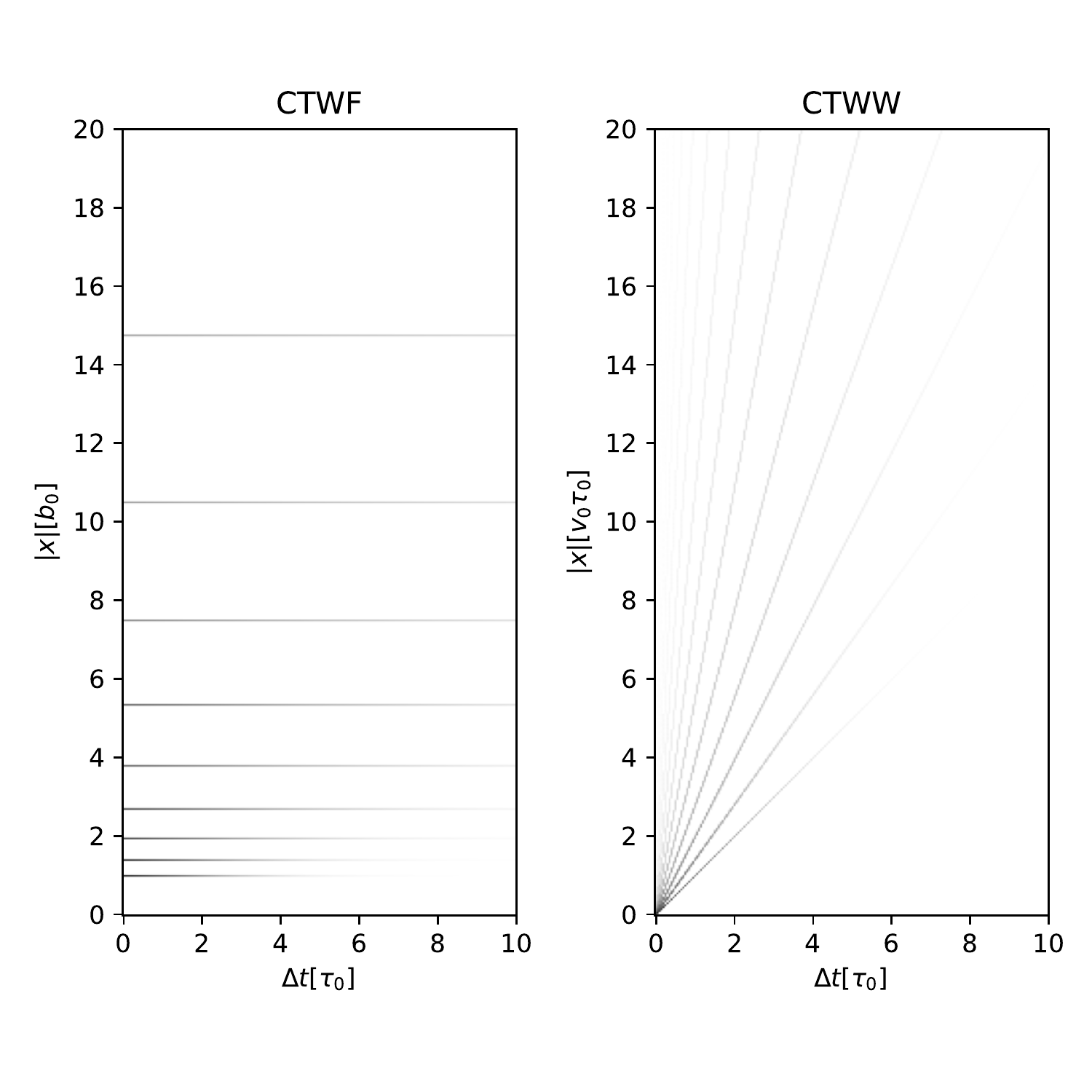}
\caption{Comparison of waiting-time and displacement distributions, WTDDs, for CTWF (left plot) and CTWW (right plot) processes based on their definitions (\ref{rown:psixDtF}) and (\ref{rown:psixDtW}), respectively. These definitions are supported herein by definition (\ref{rown:phifact}). We show heatmaps in resolution 200px by 400px and integrate WTDDs over the area of each pixel. The different shade of gray of each pixel corresponds to the different values of integrated WTDD. The brighter the grayness, the lower this value.
Because for the CTWF, displacements $\mid x\mid = b_0b^j,~j=0,1,2, \ldots $, are independent of $\Delta t$, they are shown as horizontal lines. For CTWW, analogous relationships take the form $\mid x\mid = v_0v^j\Delta t$, i.e., they linearly depend on $\Delta t,~j = 0,1,2, \ldots $. That is why we present them in form of diagonal straight lines. These horizontal and diagonal lines are numbered by index $j$ sequentially from the bottom lines. Going up on both plots we move to straight lines with increasing brightness of gray. This brightness increases also along the individual straight lines according to the exponential distribution $p_j(\Delta t)=\frac{1}{\tau _0\tau ^j}\exp\left(-\Delta t/\tau _0\tau ^j\right)$ weighted by $w(j)$ given  Eq. (\ref{rown:wj}). For both plots we assume, for example, $b= 1.4$, $v = 1.4$, $\tau = 1.3$, and $N= 1.05$.}
\label{figure:flucWFdist}
\end{center}
\end{figure}

Before we move on to the propagator's construction in Sec. \ref{section:propags24}, we clarify the general definitions of both types of processes using single-step distributions. Fig. \ref{figure:flucWFdist} is helpful here.

These are generic definitions independent of a particular form of distributions. 
They base on waiting-time and displacement joint distributions (WTDDs). 
These distributions differ both in the scaling of space variable and physical interpretations. However, we wanted this difference to be as small as possible and still leading to fundamentally different phase diagrams. In general, we can write,  
\begin{eqnarray} 
 \psi ^F(x,\Delta t)&=&\sum_{j=0}^{\infty}w(j)\frac{1}{b_0b^j}\frac{1}{\tau _0\tau ^j} \nonumber \\
 &\times &\phi \left(\frac{x}{b_0b^j},\frac{\Delta t}{\tau _0\tau ^j}\right), ~\mbox{for CTWF}
 \label{rown:psixDtF}
 \end{eqnarray}
 and 
 \begin{eqnarray} 
 \psi^W(x,\Delta t)&=&\sum_{j=0}^{\infty}w(j)\frac{1}{v_0v^j\Delta t}\frac{1}{\tau _0\tau^j} \nonumber \\
 &\times &\phi \left(\frac{x}{v_0v^j\Delta t},\frac{\Delta t}{\tau_0\tau^j}\right), ~\mbox{for CTWW} 
 \label{rown:psixDtW}
 \end{eqnarray}
 where index $j$ defines the level of the stochastic hierarchy and $j$-independent scaling function $\phi (y,\vartheta )$ is normalized that is, $\int_0^{\infty}d\vartheta\int_{-\infty}^{\infty}dy\phi(y,\vartheta)=1$. Besides, we assume that the scaling function $\phi $ is symmetric in its spatial argument $y$. We emphasize that the form of the scaling function $\phi $ is the same for CTWF and CTWW and differs only in the scaling of the spatial variable $x$. In the case of CTWF, it is $b_0b^j$, and in the case of CTWW, it is $v_0v^j\Delta t$. We continue to use indexes `F' and `W' for the CTWF and CTWW processes, respectively.
 
 What's more, we can now write conditional (or partial) WTDD assigned to the individual hierarchy level $j=0,1,2,\ldots $, in the form,
 \begin{eqnarray} 
 \psi _j^F\left(\frac{x}{b_0b^j},\frac{\Delta t}{\tau _0\tau ^j}\right)&=&\frac{1}{b_0b^j}\frac{1}{\tau _0\tau ^j} \nonumber \\
 &\times &\phi \left(\frac{x}{b_0b^j},\frac{\Delta t}{\tau _0\tau ^j}\right), ~\mbox{for CTWF}
 \label{rown:psixDtjF}
 \end{eqnarray}
and
\begin{eqnarray} 
 \psi _j^W\left(\frac{x}{v_0v^j\Delta t},\frac{\Delta t}{\tau _0\tau ^j}\right)&=&\frac{1}{v_0v^j\Delta t}\frac{1}{\tau _0\tau ^j} \nonumber \\
 &\times &\phi \left(\frac{x}{v_0v^j\Delta t},\frac{\Delta t}{\tau _0\tau ^j}\right), ~\mbox{for CTWW}, \nonumber \\
 \label{rown:psixDtjW}
 \end{eqnarray}
which makes their physical interpretation easier. 

That is, $\psi _j^F$ is the density of the conditional probability that the process is waiting at some turning point (for exactly $\Delta t$) and then executes hopping by displacement $x$ to another turning point. 
The mentioned condition means that this $\psi _j^F$ we assign to the $j^{th}$ level of the hierarchy. 
However, $\psi _j^W$, although also related to the $j^{th}$ level of the hierarchy, has a different physical interpretation. 
It is the density of the conditional probability that the process runs between successive turning points with a constant $j$-dependent velocity $v_0v^j$ precisely for the duration of $\Delta t$.

The role of index $j$ is to prioritize both displacements and the corresponding times. It is a coarse-grained approach consisting of grouping these dynamic variables according to the power of the basis $b$ and $\tau $, respectively. The power of $N$ also weights the levels of the hierarchy constructed in this way.


Notably, in the CTWW process, we are dealing with an additional coupling between single-step displacement and $\Delta t$. It is this coupling that is crucial to distinguish between the two processes discussed -- this we illustrate in Fig. \ref{figure:flucWFdist}.

\subsection{Useful relations between single-step moments}

Thanks to the definitions above, we can show what the difference between CTWW and CTWF processes is. 
We calculate for this purpose variances of the single-step displacements at the $j^{th}$ level of the hierarchy separately for both processes. 
Using Eq. (\ref{rown:psixDtjF}) we get variance at level $j$ for the CTWF in the simple form,
\begin{eqnarray} 
 \langle x^2\rangle _j^F=b_0^2b^{2j}\langle y^2\rangle ,
\label{rown:x2jF}
 \end{eqnarray}
where $\langle y^2\rangle = \int _0^{\infty }d\vartheta \int _{-\infty }^{\infty }dyy^2\phi (y,\vartheta )$ is the marginal second moment of the scale-free space variable. Similarly, by using Eq. (\ref{rown:psixDtjW}) we have,
\begin{eqnarray} 
\langle x^2\rangle _j^W=b_0^2b^{2j}\langle y^2\vartheta ^2\rangle ,
\label{rown:x2jW}
\end{eqnarray}
where $\langle y^2\vartheta ^2\rangle =\int _0^{\infty }d\vartheta \vartheta ^2\int _{-\infty }^{\infty }dyy^2\phi (y,\vartheta )$.

From Eqs. (\ref{rown:x2jF}) and (\ref{rown:x2jW}) we obtain,
\begin{eqnarray} 
\langle x^2\rangle _j^W=\langle x^2\rangle _j^F\frac{\langle y^2\vartheta ^2\rangle }{\langle y^2\rangle }.
\label{rown:x2FW}
\end{eqnarray}

For variables $x$ and $\Delta t$ separable (in a multiplicative form) for any level of the hierarchy, i.e., obeying equality $\langle y^2\vartheta ^2\rangle =\langle y^2\rangle \langle \vartheta ^2\rangle $, Eq. (\ref{rown:x2FW}) simplifies to the form, 
\begin{eqnarray} 
\langle x^2\rangle _j^W=\langle x^2\rangle _j^F\langle \vartheta ^2\rangle ,
\label{rown:x2FWsep}
\end{eqnarray}
where $\langle \vartheta ^2\rangle =\int _0^{\infty }d\vartheta \vartheta ^2\int _{-\infty }^{\infty }dy\phi (y,\vartheta )$.

In this work, we consider solely the separability considered above. 
It is clear that the difference between single-step space variances for both processes comes only from the variance of the random variable $\Delta t $ as $\langle \vartheta ^2\rangle >\langle \vartheta \rangle ^2$, i.e., the variance $\sigma _ {\vartheta}^2=\langle \vartheta ^2\rangle -\langle \vartheta \rangle ^2$ is non-vanishing. 
Of course, this variance depends in a significant way on the probability distribution $\phi $ used here to average, but it is independent of the hierarchy. 

Analogous expressions we also have for global values, i.e., averaged overall levels of the hierarchy. 
Namely,
\begin{eqnarray} 
\langle x^2\rangle ^W=\langle x^2\rangle ^F\frac{\langle y^2\vartheta ^2\rangle }{\langle y^2\rangle }=b_0^2Y\langle y^2\vartheta ^2\rangle .
\label{rown:x2FWglob}
\end{eqnarray}
corresponding to Eq. (\ref{rown:x2FW}) and
\begin{eqnarray} 
\langle x^2\rangle ^W=\langle x^2\rangle ^F\langle \vartheta ^2\rangle =b_0^2Y\langle y^2\rangle \langle \vartheta ^2\rangle ,
\label{rown:x2FWsepgl}
\end{eqnarray}
corresponding to Eq. (\ref{rown:x2FWsep}),
where $\langle x^2\rangle ^{F,W}= \sum_{j=0}^{\infty }w(j)\langle x^2\rangle _j^{F,W}$ and
\begin{eqnarray}
Y=\left\{
\begin{array}{cc}
\frac{1-\frac{1}{N}}{1-\frac{b^2}{N}}, & \mbox{for $\beta >2$,} \\
\infty , & \mbox{for $\beta <2$,}
\end{array}
\right.
\label{rown:x2FWY}
\end{eqnarray}
both for the CTWF and CTWW processes.

The first equality in (\ref{rown:x2FWsepgl}) we can rewrite in a separated additive form accessible to the physical interpretation,
\begin{eqnarray} 
\langle x^2\rangle ^W=\langle x^2\rangle ^F\sigma _{\vartheta }^2+\langle x^2\rangle ^F\langle \vartheta \rangle ^2,
\label{rown:x2FWsepglob}
\end{eqnarray}
where the first component concerns fluctuations, while the second component concerns drift.
Let us emphasize that the fluctuation component exists because the unbalanced fluctuations occur in the time symmetry breaking situation. 
That is, we assumed that the arrow of time works herein. 
As you can see the difference between CTWW and CTWF already at the level of a single displacement is significant.

We also derive equations for arbitrary even moments (stationary and non-stationary) in Appendix \ref{section:means}. 

We still assume the scaling function $\phi $ in the following simple factorized form, 
\begin{eqnarray} 
\phi(y,\theta )=\frac{1}{2}\left[\delta (y-1)+\delta (y+1)\right]
\exp(-\theta ).
\label{rown:phifact}
\end{eqnarray}
This scaling function already suffices to present the fundamental differences between CTWF and CTWW, what we show in Fig. \ref{figure:flucWFdist}. 

\subsection{Main subject}

The situation for multi-step variances is much more complicated due to the possibility of cumulative fluctuations. It leads to a significant distinction between CTWF and CTWW processes in a long time, both on the level of regular (Brownian) and singular (anomalous) behaviors. 
Explanation of this distinction is the main subject of this work. However, we must remember that the CTWW process only mimics the walk process.

We prove that CTWW provides an abundant diffusion phase diagram even in a stationary situation, consisting of several correctly classified phases. The stationary situation means here the independence of the random walk process from choosing the starting point of this process.
We show in Sec. \ref{section:msdack} that stationary diffusion phase diagrams that we build are the result of regular diffusion competition (also called linear or Brownian diffusion) with anomalous diffusion. 
What's more, each phase (even the regular one) is characterized by the autocorrelation function, which disappears according to power-law with a combined exponent. 

In the further part of the work, we compare phase diagrams for CTWW and CTWF. Besides, we calculate excess kurtosis to check if we are dealing with fractional Brownian motion (fBm) at any phase of the diagram. 
Finally, we set the explicit form of the propagator. 
We emphasize that both CTWW and CTWF are significant cases of more general CTRW formalism (see Sec. \ref{section:AppBQ} for details).

The construction and proper description of the stationary phase diagrams require the determination of moments and autocorrelation functions. To do this, we must first build a propagator $P_{st}(X,t)$. We define it as the conditional probability density of finding a walker at a position $X$ at time $t$ at the condition that the time origin can be chosen arbitrarily in the stationary situation. However, we do not require that the propagator builds in a closed-form. In this work, we consider the even moment expansion of the propagator as odd moments vanish herein. It allows us to answer the critical question in which conditions we are dealing with Gaussian walk and in which ones with non-Gaussian one.

\section{Propagator}\label{section:propags24}

\subsection{General stationary situation}

We use the following expansions of the propagator (see Appendix \ref{section:Propagatorkt} for details).
The short three-term expansion,  
\begin{eqnarray}
\tilde{P}_{st}(k,t)= 1-\frac{1}{2}k^2\langle X(t)^2\rangle _{st}+\tilde{R}_{st}^4(k,t),
\label{rown:propsdt}
\end{eqnarray}
is used for the phases where the fourth moment $\langle X(t)^4\rangle _{st}$ doesn't exist (i.e. $\langle X(t)^4\rangle _{st}=\infty $). Everywhere in this work the index $st$ is present for the stationary case. Notably, the propagator tends to the Gaussian for vanishing $k$ as then the rest $\tilde{R}_{st}^4(k,t)$ vanishes (see Eq. (\ref{rown:expans2}) in Appendix \ref{section:Propagatorkt} for details).

The longer four-term expansion, 
\begin{eqnarray}
\tilde{P}_{st}(k,t)= 1-\frac{1}{2}k^2\langle X(t)^2\rangle _{st}+\frac{1}{4!}k^4\langle X(t)^4\rangle _{st}+\tilde{R}_{st}^6(k,t), 
\nonumber \\
\label{rown:propkurt}
\end{eqnarray}
we have for phases where the fourth moment exists. In this case, we deal with Gaussian for vanishing $k$ (but greater $k$ than for the previous case) only if excess kurtosis (or equivalently the fourth cumulant) disappears that is,
\begin{eqnarray}
\kappa _{st}(t)=\langle X(t)^4\rangle _{st}-3\langle X(t)^2\rangle ^2_{st}=0
\label{rown:kurt}
\end{eqnarray}
and the rest $\tilde{R}_{st}^6(k,t)$ also vanishes (see Eq. (\ref{rown:expans4}) in Appendix \ref{section:Propagatorkt} for details). This is equivalent to the situation of vanishing cumulants of the order higher than two (if they exist). 

In general, we can write,
\begin{eqnarray}
\tilde{P}_{st}(k,t)=\sum _{j=0}^m\frac{(-1)^j}{(2j)!}k^{2j}\langle X(t)^{2j}\rangle +\tilde{R}_{st}^{2(m+1)}(k,t), 
\label{rown:prop2m}
\end{eqnarray}
for phases where the $(2m)^{th}$ moment exists (i.e. $\langle X(t)^{2m}\rangle <\infty ,~m=0,1,2,\ldots $).

Notably, in the case of anomalous diffusion at vanishing $\kappa _{st}(t)$ and $\tilde{R}^6_{st}(k,t)$, we deal with the fractional Brownian motion (fBm). However, the most interesting is the phase beyond the fBm. It includes the case of the Brownian but non-Gaussian diffusion as well as \cite{SChSPM,WKChG}. This phase splits (in the case of CTWW process) into a semi-regular diffusion phases (SRD1 and SRD2) and a regular diffusion (RD) phase (see Sec. \ref{section:msdack} and especially Fig. \ref{figure:diagramWW} for details).

\subsection{Stationary propagator and even moments: general representation}

We have (within the stationary CTRW) the following expression for the propagator \cite{KutSwit},
\begin{eqnarray}
\hat{\tilde{P}}_{st}(k,u)=\hat{\tilde{\Xi }}(k,u)+\hat{\tilde{\chi }}(k,u)\hat{\tilde{P}}(k,u)
\label{rown:propkurt1}
\end{eqnarray}
where $\hat{\tilde{F}}$ denotes the Fourier-Laplace transform of $F$ and 
\begin{eqnarray}
\hat{\tilde{P}}(k,u)=\frac{\hat{\tilde{\Psi }}(k,u)}{1-\hat{\tilde{\psi }}(k,u)}
\label{rown:propku}
\end{eqnarray}
is the propagator of the non-stationary continuous time Weierstrass walk \cite{Kutner1}. The basic quantities $\hat{\tilde{\psi }}(k,u)$, $\hat{\tilde{\Psi }}(k,u)$, $\hat{\tilde{\chi }}(k,u)$, and $\hat{\tilde{\Xi }}(k,u)$ are obtained in Appendix \ref{section:AppBQ}.

From Eqs. (\ref{rown:propkurt}), (\ref{rown:propkurt1}), and (\ref{rown:reszta}) (in Appendix \ref{section:Propagatorkt}) we get
\begin{eqnarray}
\langle X(t)^2\rangle _{st}&=&-\frac{d^2}{dk^2}\tilde{P}_{st}(k,t)|_{k=0} \nonumber \\
&\Leftrightarrow &\langle \hat{X}(u)^2\rangle _{st}=\langle \hat{X}(u)^2\rangle \nonumber \\
&-&\frac{d^2}{dk^2}\hat{\tilde{\Xi}}(k,u)|_{k=0}
-\frac{1}{u}\frac{d^2}{dk^2}\hat{\tilde{\chi}}(k,u)|_{k=0} \nonumber \\
\label{rown:X2t}
\end{eqnarray}
and
\begin{eqnarray}
\langle \hat{X}(u)^4\rangle _{st}&=&\langle \hat{X}(u)^4\rangle \nonumber \\
&+&\frac{d^4}{dk^4}\hat{\tilde{\Xi}}(k,u)|_{k=0}+\frac{1}{u}\frac{d^4}{dk^4}\hat{\tilde{\chi}}(k,u)|_{k=0}
\nonumber \\
&-&3!\langle \hat{X}(u)^2\rangle \frac{d^2}{dk^2}\hat{\tilde{\chi}}(k,u)|_{k=0}
\label{rown:X4t}
\end{eqnarray}
for the long-time limit (or $\mid u\tau  _0\mid \ll 1$), where $\tau _0$ is a time unit. Our calculations (carried out later) indicate that the last component in Eq. (\ref{rown:X4t}) does not give any contribution to $\langle \hat{X}(u)^4\rangle _{st}$ for the long-time limit.

In the definition (\ref{rown:chiDt}) of the stationary WTDD, or $\chi (t)$, there is an expected value for the non-stationary state $\langle \Delta t \rangle $. Thus, the transition to a stationary state requires the existing of this expected value. It imposes a natural limitation: there must be an exponent $\alpha> 1 $ if the stationary state is to exist. We set up converging of microscale or converging an mean waiting time in the non-stationary state. 

To better characterize the stationary diffusion phase diagram, we introduced an average waiting time in the stationary state $\langle \Delta t \rangle _ {st}$ given by Eq. (\ref{rown:tstnondt}). This average diverges for $\alpha <2 $, while converges otherwise (marginal case $\alpha = 2$ is not considered herein). Thus, the stationary diffusion phase diagram is divided into two separate parts. The first part for $\frac{1}{\alpha }>\frac{1}{2}$, where we deal with a weak ergodicity breaking \cite{Bouch,BelBar,MargBark,Heetal,RebBark,LubSokKlaf,Jeoetal,Weietal,ThieSok} and the second part for $\frac{1}{\alpha }<\frac{1}{2}$ where even full ergodicity may occur. This division takes place both for CTWF and CTWW.

\subsection{Mean-square displacement and autocorrelation function in a closed form}\label{section:msdack}

By applying the well known Mellin transform and residue techniques (as shown, for example, in our earlier works \cite{Kutner1,Kutner}) one can find the mean-square displacement (MSD) of the process for the long-time limit. It is a sum of regular ($`reg'$) and singular ($`sing'$) components,
\begin{eqnarray}
\langle X(t)^2\rangle _{st}^W=\langle X(t)^2\rangle _{st}^{Wreg}+\langle X(t)^2\rangle _{st}^{Wsing} 
\label{rown:2comps}
\end{eqnarray}
where the regular (linear or Brownian) component,
\begin{eqnarray}
\langle X(t)^2\rangle _{st}^{Wreg}=2 D_{st}^Wt,
\label{rown:regcomp}
\end{eqnarray}
always linearly depends on time $t$, with the Brownian diffusion coefficient 
\begin{eqnarray}
D_{st}^W=\frac{1}{2}\frac{\langle x^2\rangle ^W}{\langle \Delta t\rangle }, 
\label{rown:DstW}
\end{eqnarray}
where $\langle x^2\rangle ^W=\langle x^2\rangle $ is given by Eq. (\ref{rown:x2ndtm}) with $g(n=1,m=0)=2$ given by Eq. (\ref{rown:gnm}) for the CTWW, and the singular component responsible for autocorrelation,
\begin{eqnarray}
\langle X(t)^2\rangle _{st}^{Wsing}=\frac{2D_{st}^{Wf}}{\Gamma (\eta _1+1)}\left(\frac{t}{\tau _0}\right)^{\eta _1},
\label{rown:singcomp}
\end{eqnarray}
where $\eta _1\neq 1$ and the fractional diffusion coefficient 
\begin{eqnarray}
D_{st}^{Wf}=b_0^2\frac{\tau _0}{\langle{\Delta t}\rangle}\frac{1-\frac{1}{N}}{\ln N}\frac{\pi \alpha }{\mid \sin \left(\pi (\eta _1-1)\right)\mid } 
\label{rown:FracDstW}
\end{eqnarray}
with the fractional diffusion exponent 
\begin{eqnarray}
\eta _1=1+\alpha \left(\frac{2}{\beta }-1\right). 
\label{rown:Fracetaw}
\end{eqnarray}
The competition between both components in Eq. (\ref{rown:2comps}) leads to different phases depending on which component dominates at the long-time limit. Let us also take note that the horizontal red line in Fig. 2 delimiting the phases ED1 and ED2 from SRD1 and SRD2 corresponds to the situation $\eta _1=1$, which is equivalent to $\frac{1}{\beta }= \frac{1}{2}$. 

The only second component in Eq. (\ref{rown:2comps}) is taken into account whenever we calculate velocity autocorrelation function. From Eq. (\ref{rown:2comps}) we get this function in the form,
\begin{eqnarray}
C^W(t)&=&\frac{1}{2}\frac{d^2}{dt^2}\langle X(t)^2\rangle _{st}^W=\frac{1}{2}\frac{d^2}{dt^2}\langle X(t)^2\rangle _{st}^{Wsing}
\nonumber \\
&=&\frac{1}{\tau _0^2}\frac{\eta _1-1}{\Gamma (\eta _1)}D_{st}^{Wf} 
\frac{1}{\left(\frac{t}{\tau _0}\right)^{2-\eta _1}}, 
\label{rown:vaftnl}
\end{eqnarray}
where $\eta _1<2$. Keep in mind that the autocorrelation function changes from positive to negative when exponent $\eta _1$ becomes less than $1$, because in our considerations coefficient $D_{st}^{Wf}$ is positive.
Since both the CTWW and CTFW singular components are present, both corresponding autocorrelation functions are also present, decreasing for long-time limit according to power-law. As you can see, only the singular component creates the autocorrelation function with its singular behavior.

\subsection{Classification of the CTRW diffusion phases}\label{section:classification}

\begin{figure}
\begin{center}
\bigskip
\includegraphics[width=115mm,angle=0,clip]{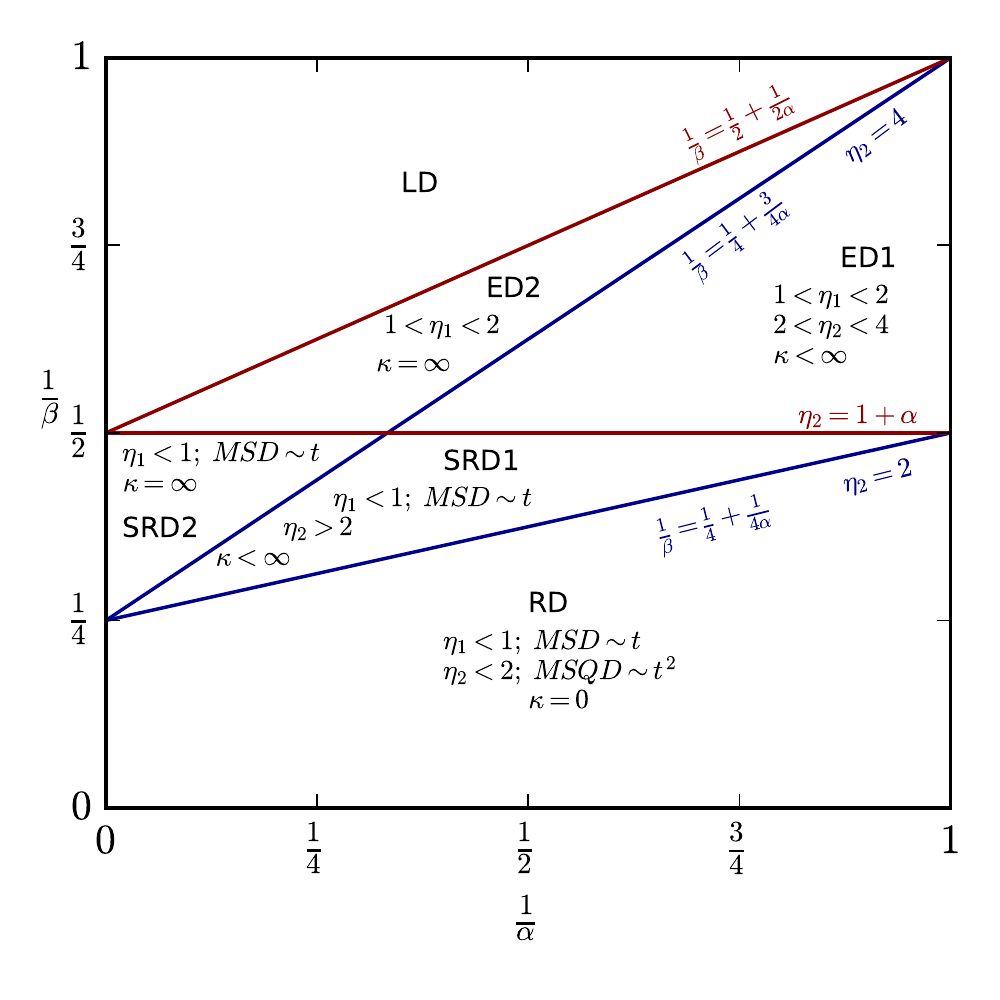}
\caption{We present the diffusion phase diagram of the fourth-order for the stationary CTWW. We show the division of the diagram into six significantly different diffusion phases. 
The detailed description of the diagram we posted in Sec. \ref{section:classification}. We remind you that we use the notations: the stationary mean-square displacement $\mbox{MSD}=\langle X^2\rangle _{st}$ and the stationary mean-square quadratic displacement $\mbox{MSQD}=\langle X^4\rangle _{st}$.}
\label{figure:diagramWW}
\end{center}
\end{figure}

In Fig. \ref{figure:diagramWW}, we presented the diffusion phase diagram based on the second and fourth moments. It is a phase diagram built on the plane determined by the partial exponents $\frac{1}{\alpha }$ and $\frac{1}{\beta }$, which govern the behavior of random walk in time and space, respectively. It contains six phases separated by straight lines (note that only four phases are marked in Fig. 1 in our earlier work \cite{KutSwit}). We discuss them in succession, going from the top of the phase diagram.

\begin{itemize}
\item[(i)] The L\'evy diffusion (LD) phase extends above the (red) sloped straight line defined by equation $\frac{1}{\beta }=\frac{1}{2}+\frac{1}{2}\frac{1}{\alpha }$. This phase is characterized by divergent MSD for finite times and in the limit of a long time. In other words, the total mean-square displacement $\langle X(t)^2\rangle _{st}^W=\infty $ for $t>0$.
\item[(ii)] The  enhanced  diffusion  (ED)  phase  is  defined  by diffusion exponent  $1<\eta _1<2$. That is, this phase  is  ruled  by  the  singular part $\langle X(t)^2\rangle _{st}^{Wsing}$, which  dominates  the regular  part $\langle X(t)^2\rangle _{st}^{Wreg}$ for long times (i.e., $t\gg \tau _0$). This phase, extending between those mentioned above (red) sloped straight line and the (red) horizontal straight line $\frac{1}{\beta }=\frac{1}{2}$, is divided into two parts marked by ED1 and ED2. These parts are separated by the segment of the oblique straight line (blue line) given by equation $\frac{1}{\beta }=\frac{1}{4}+\frac{3}{4}\frac{1}{\alpha }$, which ends at the intersection with the horizontal red line. This equation was obtained based on the analysis of the fourth moment $\langle X(t)^4\rangle _{st}$ -- we return to this point in Sec. 3.5. As you can see, three phases extend above the horizontal line $\frac{1}{\beta }=\frac{1}{2}$ and three phases below it.
\item[(iii)] The semi-regular diffusion (SRD) phase extends between this horizontal line and the oblique one (bottom black line) defined by $\frac{1}{\beta }=\frac{1}{4}+\frac{1}{4}\frac{1}{\alpha }$. For this phase $\langle X(t)^2\rangle _{st}^{Wreg}$ dominates $\langle X(t)^2\rangle _{st}^{Wsing}$ for long times. In other words, this phase characterizes by the singular diffusion exponent $\eta _1<1$ and the Brownian MSD, i.e., a linear dependence of $\langle X(t)^2\rangle _{st}^W$ on time $t$ for long times. As we show it in Sec. \ref{section:kurto} this phase is, however, the non-Gaussian one.  It is one of the most interesting results of this work. This phase we divided into parts designated by SRD1 and SRD2 through the segment of the inclined straight line given by the equation $\frac{1}{\beta }=\frac{1}{4}+\frac{3}{4}\frac{1}{\alpha }$ and starting at the intersection with the horizontal red line. Further, the fourth order characteristic of this phase we consider in Sec. \ref{section:kurto}.
\item[(iv)] Finally, below the inclined straight line (bottom black line) defined by $\frac{1}{\beta }=\frac{1}{4}+\frac{1}{4}\frac{1}{\alpha }$, extends the regular diffusion (RD) phase. This phase we define by the Brownian MSD and the singular diffusion exponent $\eta _1<1$, similarly to the SRD1 and SRD2 phases. However, in the RD phase the excess kurtosis vanishes (we discuss this matter in Sec. \ref{section:kurto}) but the other cumulants do not have to disappear. Therefore, in general, this phase should also be treated as non-Gaussian.
\end{itemize}
To sum up this part, we emphasize that all above-presented phases are (in general) of the non-Gaussian type and they are of the not-fBm. Besides, only the phase transition between LD and ED phases is discontinuous. Other phase transitions are continuous.

As you can see, each of the phases (except the LD) we characterize by a powerfully relaxing velocity autocorrelation function given by Eq. (\ref{rown:vaftnl}) with non-negative exponent $2-\eta _1$. 

It is worth to mention that the phase diagram of the stationary CTWW shown in Fig. \ref{figure:diagramWW} is a small part of the corresponding diffusion phase diagram of the non-stationary CTWW shown in Fig. 2 in our earlier work \cite{Kutner1}. It is mainly due to the limitation of the stationary CTWW to $\alpha > 1$, i.e., too much smaller fluctuations of time intervals playing a vital role in the spread of the phase diagram yet. These increased fluctuations allow you to build a phase diagram including not an only ballistic random walk, but even Richardson's turbulent random walk of passive scalars.

\subsection{Excess kurtosis}\label{section:kurto}

We set now the fourth moment or mean-square quadratic displacement (MSQD), $\langle X(t)^4\rangle _{st}$, in the same way as we did it for the second moment or MSD, $\langle X(t)^2\rangle _{st}$, using Eq. (\ref{rown:X4t}). Again, by applying the well known Mellin transform and residue techniques (as shown, for example, in our earlier works \cite{Kutner1,Kutner}) we find,
\begin{eqnarray}
\langle X(t)^4\rangle _{st}^W=\langle X(t)^4\rangle _{st}^{Wreg}+\langle X(t)^4\rangle _{st}^{Wsing}, 
\label{rown:XW4t}
\end{eqnarray}
for the long-time limit, where $Wreg$ indexes regular and $Wsing$ the singular components. We found these components in the following forms,
 \begin{eqnarray}
\langle X(t)^4\rangle _{st}^{Wreg}=12D_{st}^{W2}t^2
\label{rown:XWreg4t}
\end{eqnarray}
and
\begin{eqnarray}
\langle X(t)^4\rangle _{st}^{Wsing}=\frac{12D_{st}^{Wf2}}{\Gamma (\eta _2+1)}\left(\frac{t}{\tau _0}\right)^{\eta _2},
\label{rown:XWsing4t}
\end{eqnarray}
where 
\begin{eqnarray}
D_{st}^{W2}=\left(D_{st}^W\right)^2,~
\eta _2=1+\alpha \left(\frac{4}{\beta }-1\right), 
\label{rown:DstW2}
\end{eqnarray}
and the fractional super-Burnett coefficient 
\begin{eqnarray}
D_{st}^{Wf2}=b_0^4\frac{\tau _0}{\langle{\Delta t}\rangle}\frac{1-\frac{1}{N}}{\ln N}\frac{\pi \alpha }{\sin \left(\pi (\eta _2-1)\right)}\Pi(\eta _2), 
\label{rown:DstWf2}
\end{eqnarray}
where factor $\Pi(\eta _2)=(\eta _2-2)(\eta _2-3)$. This coefficient is positive within assumed range of $\eta _2<4$ (see Fig. \ref{figure:diagramWW} for details).

It is necessary to accept constraints imposed by the behavior of the fourth moment or excess kurtosis (blue straight sloped lines). It is required
to deeper characterize diffusion phases of the second-order (i.e., for $m=1$)  that is, driven by the finite MSD shown in Fig. \ref{figure:diagramWW} by red straight lines. It leads to six phases and not three, as would be the case for the second-order diffusion phases. 

We are now considering the properties of these additional phases by which the phase diagram on Fig. \ref{figure:diagramWW} has been enriched. More specifically, the fourth moment divides the secondary phases into parts -- we are considering these parts going sequentially from the top of the phase diagram.
\begin{itemize}
\item[(i)] Phases located above the blue sloped line given by the equation $\frac{1}{\beta }=\frac{1}{4}+\frac{3}{4}\frac{1}{\alpha }$ are characterized by $\langle X(t)^4\rangle =\infty $ or excess kurtosis $\kappa =\infty $. Above this line there are two phases: one located above the red horizontal line given by equation $\frac{1}{\beta }=\frac{1}{2}$, which from the top is limited by a sloping red line given by the equation: $\frac{1}{\beta}=\frac{1}{2}+\frac{1}{2}\frac{1}{\alpha }$, and the second phase below this horizontal line.
\item[(ii)] In the area of the acute angle between the blue sloped lines defined by the equations: the upper by the above-mentioned in item (i) and lower utilizing $\frac{1}{\beta }=\frac{1}{4}+\frac{1}{4}\frac{1}{\alpha }$. Also, in this area of the angle, there are two different phases. The phase above the horizontal long red line (mentioned above in item (i)) characterized by diffusion exponent $\eta _1>1$ and the phase below it for which $\langle X(t)^2\rangle \propto t$. 
\item[(iii)] The phase lying below (mentioned above) lower diagonal blue line is characterized by $\langle X(t)^2\rangle \propto t,~\langle X(t)^4\rangle \propto t^2,~\kappa = 0,~\eta _1<1$, and~$\eta _2<2$.
\end{itemize}
Further subdivisions of the phase diagram (shown in Fig. \ref{figure:diagramWW}) are discussed in Sec. \ref{section:extens}.

\subsection{Comparison with stationary continuous-time Weierstrass flight}\label{section:compCTWF}

For the stationary CTWW, the $1/4$ of the phase diagram surface occupies the LD phase. The finite phases occupy the remaining $3/4$ of the surface, i.e., the area for which MSD is limited for restricted times. Below in Fig. \ref{figure:diagramWF} we show that for the stationary CTWF the reverse situation takes place, i.e., the LD phase occupies $3/4$ of the phase diagram area, while the finite phases fill the remaining $1/4$ of the surface. It results in consequences that we consider below.

Technically, the analysis of the stationary CTWF is analogous to that for the stationary CTWW that's why we present our considerations only in a sketchy way. As a result, we receive equations formally identical to Eqs. (\ref{rown:2comps}) -- (\ref{rown:vaftnl}), however, solely for regions where moments for the CTWW and CTWF are simultaneously finite. That is,
\begin{eqnarray}
\langle X(t)^2\rangle _{st}^F=\langle X(t)^2\rangle _{st}^{Freg}+\langle X(t)^2\rangle _{st}^{Fsing} 
\label{rown:2compsF}
\end{eqnarray}
where
\begin{eqnarray}
\langle X(t)^2\rangle _{st}^{Freg}=\langle X(t)^2\rangle _{st}^{Wreg}
\label{rown:XFreg}
\end{eqnarray}
and
\begin{eqnarray}
\langle X(t)^2\rangle _{st}^{Fsing}=\langle X(t)^2\rangle _{st}^{Wsing}
\label{rown:regsingcompF}
\end{eqnarray}
as well
\begin{eqnarray}
C^F(t)=C^W(t).
\label{rown:CFt}
\end{eqnarray}
However, for the stationary CTWF, the regular component always dominates over the singular one. The latter component is still present, but it does not play a role for long times. The singular component yet, of course, lies at the core of the autocorrelation function of the process velocity.

We consider, for example, the phase placed below the sloped straight line defined by equation $\frac{1}{\beta }=\frac{1}{2m}-\frac{1}{2m}\frac{1}{\alpha }$, where $m=1,2,\ldots $ (cf. Fig. \ref{figure:diagramWF}). The number $2m$ defines, herein, the order of the phase, where above this line or threshold the moment of the $(2m)^{th}$ order diverges but below it, it converges. In this sense, we can say that we deal with discontinuous phase transitions. As you can see, the phases are located in a descending order: the higher-order phase is contained within the lower order phase. We are dealing with a phase diagram of $2m$ order (i.e., containing $m+1$ phases) if we no longer divide the phase of the order of $2m$ into higher-order phases. Thus we get a tool that allows you to distinguish phases of any orders. Let's also notice that the areas of finite diffusion phases (i.e., the phases placed below the borderline defined by equation $\frac{1}{\beta }=\frac{1}{2}-\frac{1}{2}\frac{1}{\alpha }$) are now strongly reduced. That is, the stationary CTWF has a diffusion phase diagram, which is a small part (i.e., one third) of the CTWW phase diagram (see Figs. \ref{figure:diagramWW} and \ref{figure:diagramWF} for the comparison).

Let us consider, for example, the phase of the $4^{th}$ order. There are only three diffusion phases shown in Fig. \ref{figure:diagramWF}. The one above the sloped straight line given by the equation $\frac{1}{\beta }=\frac{1}{2}-\frac{1}{2}\frac{1}{\alpha }$ is characterized by divergent MSD. The second phase placed below this line is the SRD2 phase, for which MSD depends in a Brownian (linear) manner from time for long times, while having kurtosis diverging. The third phase placed below this line $\frac{1}{\beta }=\frac{1}{4}-\frac{1}{4}\frac{1}{\alpha }$, is characterized by vanishing  kurtosis. However, the higher cumulants in this phase exist. That's why this phase, although it is Brownian, it is a non-Gaussian one. 
 
\begin{figure}
\begin{center}
\bigskip
\includegraphics[width=110mm,angle=0,clip]{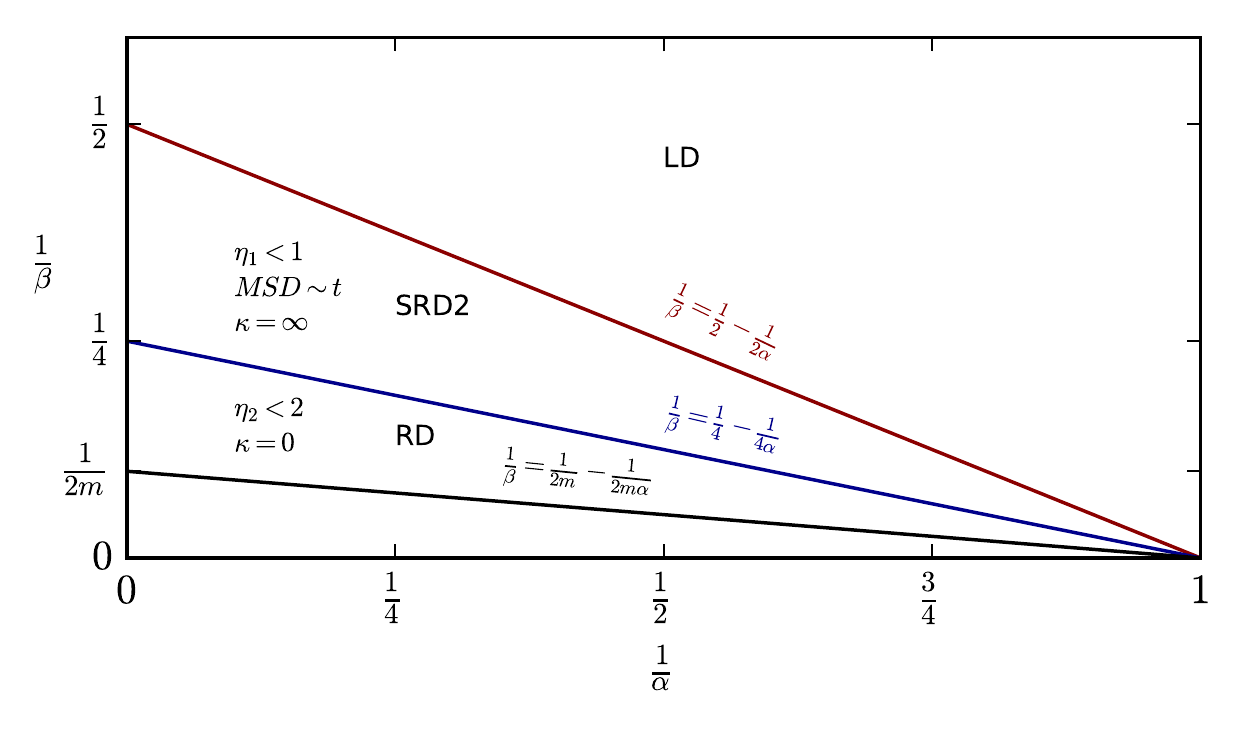}
\caption{The diffusion phase diagram for the stationary CTWF. The plot also has a borderline $\frac{1}{\beta }=\frac{1}{2m}-\frac{1}{2m}\frac{1}{\alpha }$ for any $m$ value that allows you to build a phase diagram of the order of $2m$. Analogous boundary lines for CTWW we show in Fig. \ref{figure:dpd2m}. We divided the diagram by the borderlines of this type for $m=1$ and $m=2$ into significantly different diffusion phases -- the detailed description of the diagram we have in the main text.}
\label{figure:diagramWF}
\end{center}
\end{figure}

\section{Further extension of the CTWW diffusion phase diagram}\label{section:extens}

We continue to extend the results shown above for the CTWW to any even moments $\langle X(t)^{2n}\rangle $. This extension is technically more complicated than the corresponding extension for the CTWF discussed in Sec. \ref{section:compCTWF} due to the more fluctuations present in the CTWW.

To determine any even moment of the process $X(t)$ for both CTWW and CTWF in a closed form for long times, we use the Mellin transform technique (which we outline in Appendix \ref{section:auxil}). This approach allows to obtain finite sum values in expressions (\ref{rown:XiWku}) and (\ref{rown:XiFku}) in Appendix \ref{section:auxil} provided that inequality occurs,
\begin{eqnarray}
\frac{1}{\beta }<\frac{1}{2m}+\frac{c-2}{2m}\frac{1}{\alpha },~m=1,2,\ldots ,
\label{rown:2inequal}
\end{eqnarray}
where $c=2m+1$ for the CTWW and $c=1$ for the CTWF.

We can check that inequality (\ref{rown:2inequal}) provides the most reliable condition, i.e., the straight boundary line defined by equality $\frac{1}{\beta }=\frac{1}{2m}+\frac{c-2}{2m}\frac{1}{\alpha }$ is the lowest on the phase diagram compared to the analogous boundary straight lines provided by the conditions imposed by the other components of MSD.

Moreover, the conjecture is assumed, that the exponent governing the singular component, $\langle X(t)^{2m}\rangle _{st}^{Wsing}$, of the $(2m)^{th}$ order's moment, $\langle X(t)^{2m}\rangle _{st}$ takes the generic form, 
\begin{eqnarray}
\eta _m=1+\alpha \left(\frac{2m}{\beta }-1\right),~m=1,2,\ldots ,
\label{rown:etam}
\end{eqnarray} 
where separation,
\begin{eqnarray}
\langle X(t)^{2m}\rangle _{st}=\langle X(t)^{2m}\rangle _{st}^{Wreg}+\langle X(t)^{2m}\rangle _{st}^{Wsing}
\label{rown:separ2m}
\end{eqnarray} 
was assumed. Of course, the formula (\ref{rown:etam}) is a generalization of those previously derived for $m=1$ Eq. (\ref{rown:singcomp}) and $m=2$ Eq. (\ref{rown:XWsing4t}). Analogously, expression (\ref{rown:separ2m}) is a generalization of Eqs. (\ref{rown:2comps}) and (\ref{rown:XW4t}), respectively.

\begin{figure}
\begin{center}
\bigskip
\includegraphics[width=110mm,angle=0,clip]{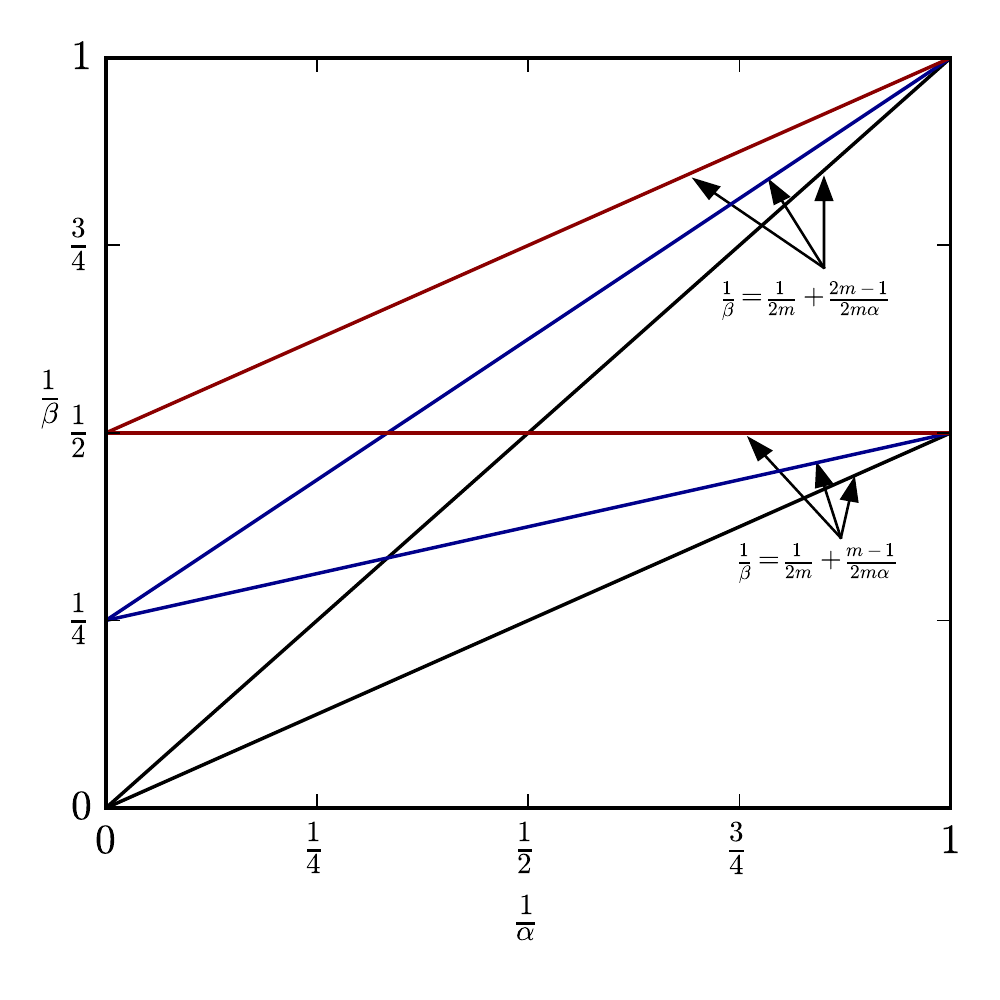}
\caption{The diffusion phase diagram of the $2m^{th}$ order for the stationary CTWW. The arms of the acute angle with the apex of the angle at the point $\left(0,\frac{1}{2m}\right)$ determine the dominant domain of the singular component $\langle X(t)^{2m}\rangle _{st}^{Wsing}$ of the order of $2m$. The arms of the angle are described by equations, respectively: the upper by $\frac{1}{\beta }=\frac{1}{2m}+\frac{2m-1}{2m}\frac{1}{\alpha }$ and lower one by $\frac{1}{\beta }=\frac{1}{2m}+\frac{m-1}{2m}\frac{1}{\alpha }$.} 
\label{figure:dpd2m}
\end{center}
\end{figure}

We analyze the consequences of the formula (\ref{rown:etam}).
\begin{itemize}
\item[(i)] The area of the phase diagram defined by inequality $\eta _m>m$ refers to the case when the singular component in the formula (\ref{rown:separ2m}) dominates over the regular component. This inequality is equivalent to the following one, $\frac{1}{\beta }>\frac{1}{2m}+\frac{m-1}{2m}\frac{1}{\alpha }$. Hence, equality $\frac{1}{\beta }=\frac{1}{2m}+\frac{m-1}{2m}\frac{1}{\alpha }$ defines the border line below which is the area of the phase diagram (marked by SDR1 and SDR2) where the regular component in Eq. (\ref{rown:separ2m}) dominates the singular component. In Fig. \ref{figure:diagramWW}, this equation has solutions represented by the bottom red and bottom blue lines for $m=1$ and $m=2$, respectively. In Fig. \ref{figure:dpd2m} we have additionally presented a solution for $m=\infty $ (bottom black line).  
\item[(ii)] Sharp angles bounded by sides that are given above in (i) by the equality and earlier introduced equation $\frac{1}{\beta }=\frac{1}{2m}+\frac{2m-1}{2m}\frac{1}{\alpha }$, define the diffusion phases.
\end{itemize}

\section{Concluding remarks}

This work is an extension of our previous one \cite{KutSwit}, especially when it comes to in-depth analysis of diffusion phases. The main focus of this work is the continuous-time Weierstrass walk, which contains more fluctuations than the continuous-time Weierstrass flight. It is because trajectories of the CTWW process fluctuate around the corresponding trajectories of the CTWF process (see Fig. \ref{figure:flucWF} for details). As you can see, fluctuations can refer to every turning point at every level of the CTWW hierarchy -- we can call it `the hierarchy of partial fluctuations'. Any of such a partial fluctuation is directly related to the relationship (\ref{rown:propWtermY}) in Appendix \ref{section:propagator}.

In this work, we compared the continuous-time Weierstrass walk with continuous-time Weierstrass flight and discussed the most significant consequences flowing here, for example, containing the diffusive phase diagrams' comparison. These two stochastic processes provide alternative descriptions of random motion in continuous time (see Figs. \ref{figure:flucWF} and \ref{figure:flucWFdist} for details). We showed both processes have a common area of diffusion diagram.  There, these formalisms are identical. There is also much more extensive area where they are not (see Figs. \ref{figure:diagramWW} and \ref{figure:diagramWF} for details). We observed that CTWW formalism is abundant and more flexible than CTWF. It is because the CTWF goes into the L\'evy diffusion (i.e., into the divergent of the process MSD) on a much larger area of the diffusion phase diagram than the CTWW one.

The added value of this work is also the analysis of higher moments (see Fig. \ref{figure:dpd2m} for details). It is essential in the study of higher-order phase transitions as well as in the expansion of CTWW and CTWF processes on multifractal stochastic processes \cite{PMKK,KKPM}.

In conclusion, in this work, we showed how dynamic random variables' coupling on each level of hierarchical random walk given by Eq. (\ref{rown:propWtermY}) influences on both anomalous and Brownian but mainly non-Gaussian diffusion in the frame of the stationary continuous-time Weierstrass walk. The most important consequence of this influence is the existence of powerfully relaxing autocorrelation functions controlled by the singular exponent $\eta _1$ (see Eqs. (\ref{rown:singcomp}), (\ref{rown:vaftnl}), and (\ref{rown:CFt})). The final long-term behavior of the system is the result of competition between both worlds: regular and singular. This type of approach allows you to model time series for different time horizons from the real world. For example, from financial markets, where daily time series consist of alternating stationary and non-stationary fragments -- the latter is, for example, growing and bursting stock bubbles. 
Our approach can be extended to the non-stationary CTWW process -- the first step in this direction we made at works \cite{Kutner1,Kutner2}.

\appendix 

\section{Expansions of propagator}\label{section:Propagatorkt}

We consider expansions of the Fourier transform (characteristic function) of the propagator,
\begin{eqnarray}
\tilde{P}_{st}(k,t)=\int_{-\infty}^{\infty}dX\exp(ikX)P_{st}(X,t)
\label{rown:expans}
\end{eqnarray} 
for three cases. 

\begin{itemize}
\item[(i)] For the phases where the fourth moment diverges (doesn't exist), we have from Eq. (\ref{rown:expans}) the expansion,
\begin{eqnarray}
\tilde{P}_{st}(k,t)&=&1-\frac{1}{2}k^2\int_{-\infty}^{\infty}dX X^2P_{st}(X,t) \nonumber \\
&+&\int_{-\infty}^{\infty}dX\sum_{n=2}^{\infty }\frac{(-1)^n}{(2n)!}k^{2n}X^{2n}P_{st}(X,t). \nonumber \\
\label{rown:expans2}
\end{eqnarray} 
\item[(ii)] For the phases where the second and fourth moments exist, we derive from Eq. (\ref{rown:expans}) the expansion,
\begin{eqnarray}
\tilde{P}_{st}(k,t)&=&1-\frac{1}{2}k^2\int_{-\infty}^{\infty}dX X^2P_{st}(X,t) \nonumber \\
&+&\frac{1}{4!}k^4\int_{-\infty}^{\infty}dX X^4P_{st}(X,t) \nonumber \\
&+&\int_{-\infty}^{\infty}dX\sum_{n=3}^{\infty }\frac{(-1)^n}{(2n)!}k^{2n}X^{2n}P_{st}(X,t). \nonumber \\
\label{rown:expans4}
\end{eqnarray} 
\item[(iii)] In general, when the moment of the order of $2m$ exists, we receive,
\begin{eqnarray}
\tilde{P}_{st}(k,t)&=&\sum _{j=0}^m\frac{(-1)^j}{(2j)!}k^{2j}\int_{-\infty}^{\infty}dX X^{2j}P_{st}(X,t) \nonumber \\
&+&R_{st}^{2(m+1)}(k,t),
\label{rown:expans2m}
\end{eqnarray}
where
\begin{eqnarray}
R_{st}^{2(m+1)}(k,t)&=& \nonumber \\
&=&\int_{-\infty}^{\infty}dX\sum_{n=m+1}^{\infty }\frac{(-1)^n}{(2n)!}k^{2n}X^{2n}P_{st}(X,t);
\label{rown:expansR2m}
\end{eqnarray}
we remind that all odd moments vanish.
\end{itemize}

In all above formulas (\ref{rown:expans2}), (\ref{rown:expans4}), and (\ref{rown:expans2m}) it is not possible to change the integration with summation in last components and to integrate the term after the term one by one, because the obtained integrals (moments of variable X) are divergent for $n>m$. Moreover, the integral of the sum of the odd terms disappeared, because the propagator $P(X,t)$ is an even function of the variable $X$ (no drift is present in both random walks).

As for the notations, the second ingredient in Eqs. (\ref{rown:expans2}) and (\ref{rown:expans4}) is the second moment $\langle X(t)^2\rangle $, the third ingredient in Eq. (\ref{rown:expans4}) is the fourth moment $\langle X(t)^4\rangle $, while the last ingredient in both equations is the rest $\tilde{R}_{st}^{2(n+1)}(k,t)$ of the expansions, where $n = 1$ for Eq. (\ref{rown:expans2}), $n = 2$ for Eq. (\ref{rown:expans4}), and in general $n=m+1$ for Eq. (\ref{rown:expans2m}). Indeed, this notation is used in Sec. \ref{section:propags24}.

You also need to pay attention to the important property of the rest defined by the second component in Eq. (\ref{rown:expans2m}),
\begin{eqnarray}
\frac{d^{2m+1}}{dk^{2m+1}}\tilde{R}_{st}^{2m+1}(k,t)|_{k=0}=0.
\label{rown:reszta}
\end{eqnarray}
It results directly from the definition of the rest, where variable $k^{2(m+1)}$ can be pulled out in front of the integral (see Eqs. (\ref{rown:expans2}) and (\ref{rown:expans4}) for characteristic examples).

\subsection{Exact stationary CTWW and CTWF propagators}

\subsubsection{CTWW Propagator}\label{section:propagator}

According to generic Eq. (\ref{rown:propkurt1}), specific expression for the stationary propagator $\hat{\tilde{P}}_{st}^W(k,u)$ of the CTWW takes the form,
\begin{eqnarray}
\hat{\tilde{P}}_{st}^W(k,u)=\hat{\tilde{\Xi }}^W(k,u)+\hat{\tilde{\chi }}^W(k,u)\hat{\tilde{P}}^W(k,u),
\label{rown:propkurW}
\end{eqnarray}
where
\begin{eqnarray}
\hat{\tilde{\Xi }}^W(k,u)&=&\tau _0\left(1-\frac{\tau }{N}\right)\sum _{j=0}^{\infty }\left(\frac{\tau ^2}{N}\right)^j
\hat{\tilde{\Theta }}_j^W(k,u), \nonumber \\
\hat{\tilde{\chi }}^W(k,u)&=&\left(1-\frac{\tau }{N}\right)\sum _{j=0}^{\infty }\left(\frac{\tau }{N}\right)^j\hat{\tilde{\Theta }}_j^W(k,u), \nonumber \\
\hat{\tilde{P}}^W(k,u)&=&\frac{\hat{\tilde{\Psi}}^W(k,u)}{1-\hat{\tilde{\psi}}^W(k,u)} \nonumber \\
&=&\tau _0\frac{\sum_{j=0}^{\infty }\left(\frac{\tau }{N}\right)^j\hat{\tilde{\Theta }}_j^W(k,u)}{\sum_{j=0}^{\infty }\left(\frac{1}{N}\right)^j\left(1-\hat{\tilde{\Theta }}_j^W(k,u)\right)},
\label{rown:propWterm}
\end{eqnarray}
with
\begin{eqnarray}
\hat{\tilde{\Theta }}_j^W(k,u)=\frac{u\tau _0\tau ^j+1}{(u\tau _0\tau ^j+1)^2+k^2(b_0b^j)^2}.
\label{rown:propWtermX}
\end{eqnarray}
and
\begin{eqnarray}
\hat{\tilde{\psi}}^W(k,u)&=&\left(1-\frac{1}{N}\right)\sum_{j=0}^{\infty }\left(\frac{1}{N}\right)^j\hat{\tilde{\Theta }}_j^W(k,u), 
\nonumber \\
\hat{\tilde{\Psi}}^W(k,u)&=&\tau _0\left(1-\frac{1}{N}\right)\sum_{j=0}^{\infty }\left(\frac{\tau }{N}\right)^j\hat{\tilde{\Theta }}_j^W(k,u).
\label{rown:propWtermpP}
\end{eqnarray}

The form of Eq. (\ref{rown:propWtermX}) is the result of a hierarchical spatial-temporal coupling and a simple dynamic coupling between $x$ and $\Delta t$ at each level $j$ of the hierarchy, 
\begin{eqnarray}
x(\Delta t)=\pm v_0v^j\Delta t
\label{rown:propWtermY}
\end{eqnarray}
resulting from $\delta $-s present in the definition (\ref{rown:phifact}), where $b_0=v_0\tau _0$ and $b=v\tau $. Each sign in Eq. (\ref{rown:propWtermY}) is drawn with equal probability $1/2$, because there is no drift in the system.

Let us emphasize that Eq. (\ref{rown:propWtermY}) is the equation of stochastic dynamics without drift, where the single-step displacement $x(\Delta t)=X (t + \Delta t) -X (t)$, the probability of its orientation $\pm $ is $prob(+)=prob(-)=0.5$, its random discrete variable (or index) $j$ comes from the geometric distribution $w(j)=\left(1-\frac{1}{N}\right)\frac{1}{N^j}$ and the single-step inter-event time $\Delta t$ from the exponential distribution $p_j(\Delta t)=\frac{1}{\tau _0\tau ^j}\exp\left(-\Delta t/\tau _0\tau ^j\right)$. As you can see, the stochasticity of this dynamics is governed by three random variables drawn from three different probability distributions. What's more, it is the equation of stochastic dynamics with a random time step.

Notably, for the CTWF the analogous equation of stochastic dynamics takes a much simpler form,
\begin{eqnarray}
x(\Delta t)=\pm b_0b^j,
\label{rown:propFtermY}
\end{eqnarray}
where $b_0=v_0\tau _0$ and $b=v\tau $.

Both stationary and non-stationary CTWW are characterized by: (i) the random nature of $\Delta t$ variable, as this variable is subject to the exponential distribution $p_j(\Delta t)$ and (ii) time-space coupling (\ref{rown:propWtermY}) at any level of the walk's hierarchy, although variables $x$ and $\Delta t$ are uncorrelated, (iii) the hierarchical nature of this walk. To see the uncorrelation mentioned above, just write the cross-correlation function in the form $\langle x\Delta t\rangle - \langle x\rangle \langle \Delta t\rangle =\langle v_0v^j\rangle \left[\langle (\Delta t)^2\rangle - \langle \Delta t\rangle ^2\right]=0$ as a dichotomic averaging $\langle v_0v^j\rangle =+\frac{1}{2}v_0v^j-\frac{1}{2}v_0v^j=0$ (no drift is present in the system). 

The list of items mentioned above has consequences that are the subject of this work. We examine these consequences in comparison with the corresponding ones of the CTWF model, for which dependence (\ref{rown:propWtermY}) is replaced by Eq. (\ref{rown:propFtermY}). For the CTWF, $x$ and $\Delta t$ are also uncorrelated. 

For both models any even functions of $x$ are correlated with any functions of $\Delta t$ but giving different values (see Appendix \ref{section:means} for details).

Note that equality occurs
\begin{eqnarray}
\hat{\tilde{\Theta }}_j^W(k=0,u)=\frac{1}{u\tau _0\tau ^j+1},
\label{rown:propWtermTh}
\end{eqnarray}
which is very useful below. Hence, and from the last equality in (\ref{rown:propWterm}) we get,
\begin{eqnarray}
\hat{\tilde{P}}^W(k=0,u)=\frac{1}{u}.
\label{rown:propWtermP0}
\end{eqnarray}

\subsubsection{CTWF Propagator}

For the CTWF, analogical relations take place. Namely, Eq. (\ref{rown:propkurW}) takes the form,
\begin{eqnarray}
\hat{\tilde{P}}_{st}^F(k,u)=\hat{\tilde{\Xi }}^F(k,u)+\hat{\tilde{\chi }}^F(k,u)\hat{\tilde{P}}^F(k,u),
\label{rown:propkurF}
\end{eqnarray}
where
\begin{eqnarray}
\hat{\tilde{\Xi }}^F(k,u)&=&\tau _0\left(1-\frac{\tau }{N}\right)\sum _{j=0}^{\infty }\left(\frac{\tau ^2}{N}\right)^j
\hat{\tilde{\Theta }}_j^F(k,u), \nonumber \\
\hat{\tilde{\chi }}^F(k,u)&=&\left(1-\frac{\tau }{N}\right)\sum _{j=0}^{\infty }\left(\frac{\tau }{N}\right)^j\hat{\tilde{\Theta }}_j^F(k,u), \nonumber \\
\hat{\tilde{P}}^F(k,u)&=&\frac{\hat{\tilde{\Psi}}^F(k=0,u)}{1-\hat{\tilde{\psi}}^F(k,u)} \nonumber \\
&=&\tau _0\frac{\sum _{j=0}^{\infty }\left(\frac{\tau }{N}\right)^j\hat{\tilde{\Theta }}_j^F(k=0,u)}{\sum_{j=0}^{\infty }\left(\frac{1}{N}\right)^j\left(1-\hat{\tilde{\Theta }}_j^F(k,u)\right)},
\label{rown:propFterm}
\end{eqnarray}
with
\begin{eqnarray}
\hat{\tilde{\Theta }}_j^F(k,u)=\cos (kb_0b^j)\hat{\tilde{\Theta }}_j^W(k=0,u)
\label{rown:propFtermX}
\end{eqnarray}
and
\begin{eqnarray}
\hat{\tilde{\psi}}^F(k,u)&=&\left(1-\frac{1}{N}\right)\sum _{j=0}^{\infty }\left(\frac{1}{N}\right)^j\hat{\tilde{\Theta }}_j^F(k,u), 
\nonumber \\
\hat{\tilde{\Psi}}^F(k,u)&=&\tau _0\left(1-\frac{1}{N}\right)\sum _{j=0}^{\infty }\left(\frac{\tau }{N}\right)^j\hat{\tilde{\Theta }}_j^F(k,u),
\label{rown:propFtermpP}
\end{eqnarray}
which are simpler as Eq. (\ref{rown:propFtermX}) has simpler form than the corresponding Eq. (\ref{rown:propWtermX}).

Thanks to Eq. (\ref{rown:propFtermX}), we obtain many useful properties regarding marginal distributions. Namely, 
\begin{eqnarray}
\hat{\tilde{\Theta }}_j^F(k=0,u)&=&\hat{\tilde{\Theta }}_j^W(k=0,u), \nonumber \\
\hat{\tilde{\psi }}^F(k=0,u)&=&\hat{\tilde{\psi }}^W(k=0,u), \nonumber \\
\hat{\tilde{\Psi }}^F(k=0,u)&=&\hat{\tilde{\Psi }}^W(k=0,u), \nonumber \\
\hat{\tilde{P}}^F(k=0,u)&=&\hat{\tilde{P}}^W(k=0,u), \nonumber \\
\hat{\tilde{\Xi }}^F(k=0,u)&=&\hat{\tilde{\Xi }}^W(k=0,u), \nonumber \\
\hat{\tilde{\chi }}^F(k=0,u)&=&\hat{\tilde{\chi }}^W(k=0,u), \nonumber \\
\hat{\tilde{P}}_{st}^F(k=0,u)&=&\hat{\tilde{P}}_{st}^W(k=0,u), \nonumber \\
\frac{d^{2m+1}}{dk^{2m+1}}\hat{\tilde{\Theta }}_j^F(k,u)\mid _{k=0}&=&\frac{d^{2m+1}}{dk^{2m+1}}\hat{\tilde{\Theta }}_j^W(k,u)\mid _{k=0}=0, \nonumber \\
\frac{d^{2m}}{dk^{2m}}\hat{\tilde{\Theta }}_j^W(k,u)\mid _{k=0}&=&(-1)^m 2m!(b_0)^{2m}b^{2mj} \nonumber \\
&\times &\left[\hat{\tilde{\Theta }}_j^W(k=0,u)\right]^{2m+1}, \nonumber \\
\frac{d^{2m}}{dk^{2m}}\hat{\tilde{\Theta }}_j^F(k,u)\mid _{k=0}&=&(-1)^m(b_0)^{m}b^{2mj}\hat{\tilde{\Theta }}_j^F(k=0,u). \nonumber \\
\label{rown:propFXYZ}
\end{eqnarray}

\subsection{Auxiliary calculations}\label{section:auxil}

We carry out calculations that form the basis for obtaining any expression containing an infinite summation after the current variable $j$ in a closed-form. For example, a typical case is
\begin{eqnarray}
& &\frac{d^{2m}}{dk^{2m}}\hat{\tilde{\Xi }}^W(k,u)\mid _{k=0} \nonumber \\
&=&\tau _0\left(1-\frac{\tau }{N}\right)\sum _{j=0}^{\infty }\left(\frac{\tau ^2}{N}\right)^j
\frac{d^{2m}}{dk^{2m}}\hat{\tilde{\Theta }}_j^W(k,u)\mid _{k=0} \nonumber \\
&=&(-1)^m 2m!(b_0)^{2m}\tau _0\left(1-\frac{\tau }{N}\right)\sum _{j=0}^{\infty }\left(\frac{\tau ^2b^{2m}}{N}\right)^j \nonumber \\
&\times &\left(\frac{1}{u\tau _0\tau ^j+1}\right)^{2m+1},~m=1,2,\ldots .
\label{rown:sumjMel}
\end{eqnarray}
To perform summation we replace $\left(\frac{1}{u\tau _0\tau ^j+1}\right)^{2m+1}$ by its inverse Mellin transformation so that instead of Eq. (\ref{rown:sumjMel}) we can write,
\begin{eqnarray}
& &\frac{d^{2m}}{dk^{2m}}\hat{\tilde{\Xi }}^W(k,u)\mid _{k=0} \nonumber \\
&=&(-1)^m 2m!(b_0)^{2m}\tau _0\left(1-\frac{\tau }{N}\right)\sum _{j=0}^{\infty }\left(\frac{\tau ^2b^{2m}}{N}\right)^j \nonumber \\
&\times &\frac{1}{2\pi i}\int _{c-i\infty }^{c+i\infty }ds\tau ^{-sj}\frac{\pi }{\sin(\pi s)}(u\tau _0)^{-s}
\left(\begin{array}{c}s-1\\2m\end{array}\right),
\label{rown:sumjMelm}
\end{eqnarray}
where $0<c=\Re ~s<2m+1$ and $\left(\begin{array}{c}s-1\\2m\end{array}\right)$ is the generalized binomial (Newton) coefficient.

The sum and integral can be interchanged in Eq. (\ref{rown:sumjMelm}) only when they converge. After such a change, we are dealing with a geometric series, $\left(\frac{b^{2m}}{\tau ^{s-2}N}\right)^j$, which is convergent if and only if its quotient $\frac{b^{2m}}{\tau ^{c-2}N} <1$. This inequality is equivalent (to very good approximation) to the following the essential one,
\begin{eqnarray}
\frac{1}{\beta }<\frac{1}{2m}+\frac{2m-1}{2m}\frac{1}{\alpha },
\label{rown:sumjMelm2}
\end{eqnarray}
where the upper limit $c=2m+1$ was used.

Thus, if constrain (\ref{rown:sumjMelm2}) is satisfied then Eq. (\ref{rown:sumjMelm}) takes the form,
\begin{eqnarray}
\frac{d^{2m}}{dk^{2m}}\hat{\tilde{\Xi }}^W(k,u)\mid _{k=0}=(-1)^m (2m)!(b_0)^{2m}\tau _0\left(1-\frac{\tau }{N}\right) \nonumber \\
\times \frac{1}{2\pi i}\int _{c-i\infty }^{c+i\infty }ds\frac{1}{1-\frac{b^{2m}}{\tau ^{s-2}N}}\frac{\pi }{\sin(\pi s)}
\left(\begin{array}{c}s-1\\2m\end{array}\right)(u\tau _0)^{-s}. \nonumber \\
\label{rown:sumjMelm3}
\end{eqnarray}
This form exhibits the origin of singularities coming from the hierarchical structure of the process random walk which are clustered accrording to the geometric series. The integrand has poles of the first order from $\frac{1}{\sin(\pi s)}$ at $s=s_0=0,-1,-2,\ldots ,$ and from 
$\frac{1}{1-\frac{b^{2m}}{\tau ^{s-2}N}}$ at $s=s_1(n)=2+\alpha \left(\frac{2m}{\beta }-1\right)\pm 2\pi i\frac{n}{\ln \tau },~n=0,1,\ldots $
Integration is performed using the well-known residual method, as it is possible to build a contour (here rectangular) encompassing these poles. The poles $s_0=0$ and $s_1(n = 0)$ are still selected, because the remaining ones give a vanishing contribution to the integral (\ref{rown:sumjMelm3}) for $\mid u\mid \ll 1$, which we show below. The right side of the contour is parallel to the vertical imaginary axis and has a real coordinate just equal to $c$. It shows that the integral on the other sides of the contour disappear as they move away to $\pm \infty $. Hence, Eq. (\ref{rown:sumjMelm3}) takes the form,
\begin{eqnarray}
\frac{d^{2m}}{dk^{2m}}\hat{\tilde{\Xi }}^W(k,u)\mid _{k=0}=(-1)^m (2m)!(b_0)^{2m}\tau _0\left(1-\frac{\tau }{N}\right) \nonumber \\
\times \left[\frac{1}{1-\frac{\tau ^2b^{2m}}{N}}+\frac{1}{\ln N}\frac{\pi \alpha }{\sin(\pi s_1)}\left(\begin{array}{c}s_1-1\\2m\end{array}\right)(u\tau _0)^{-s_1}\right] \nonumber \\
\label{rown:sumjMelm3fin}
\end{eqnarray}
where simplified notation $s_1(n=0)=s_1$ is used.

\subsection{The key marginal equalities}\label{section:margequal}

We present a source of key equations that define the upper boundaries of the $\mbox{MSD} <\infty $ areas on CTWW and CTWF phase diagrams. It can be shown that these upper boundaries are dictated by the behavior of $\hat{\tilde{\Xi }}(k,u)$ -- its appropriate derivatives are present in the definition of any even moments of the process $X(t)$ (cf. Eq. (\ref{rown:propkurt1}) and Eqs. (\ref{rown:X2t}) and (\ref{rown:X4t}) for example). For the CTWW we have,
\begin{eqnarray}
\frac{d^{2m}}{dk^{2m}}\hat{\tilde{\Xi }}^W(k,u)\mid _{k=0}=(-1)^m 2m!(b_0)^{2m}\tau _0\left(1-\frac{\tau }{N}\right) \nonumber \\
\times \sum _{j=0}^{\infty }\left(\frac{\tau ^2b^{2m}}{N}\right)^j\left[\hat{\tilde{\Theta }}_j^W(k=0,u)\right]^{2m+1},
\label{rown:XiWku}
\end{eqnarray}
while for the CTWF we get simpler form,
\begin{eqnarray}
\frac{d^{2m}}{dk^{2m}}\hat{\tilde{\Xi }}^F(k,u)\mid _{k=0}=\tau _0\left(1-\frac{\tau }{N}\right) \nonumber \\
\times \sum _{j=0}^{\infty }\left(\frac{\tau ^2b^{2m}}{N}\right)^j\hat{\tilde{\Theta }}_j^F(k=0,u).
\label{rown:XiFku}
\end{eqnarray}

The fact that $\hat{\tilde{\Theta }}_j^F(k=0,u)$ for CTWF occurs in the first power independently of $m$ makes an essential difference between CTWF and CTWW key. The existence of $\hat{\tilde{\Theta }}_j^W(k=0,u)$ in power $m$ for CTWW is the result of a simple spatial-temporal coupling (\ref{rown:propWtermY}) at each level $j$ of the random walk hierarchy. For the CTWF such a coupling is absent.

\section{Basic distributions and means}\label{section:AppBQ}

\subsection{Basic distributions}

The WTDD $\psi (x,\Delta t)$ is the primary distribution. The non-stationary and stationary CTRW formalisms (including CTWW and CTWF ones, which are typical hierarchical cases) we based on it. It is the probability density of the walker single-step displacement $x$ in its duration time $\Delta t$. After this time the walker begins its next single step by marking the turning point of its trajectory. In both our cases (defined by walks and flights) WTDD represents in the form of the Weierstrass distributions.

Our approach is two-step: first, we define WTDDs for the non-stationary CTWF and CTWW, and then we make them stationary. Namely, we have
\begin{eqnarray}
\psi (x,\Delta t)&=&\frac{1}{2}\left(1-\frac{1}{N}\right)
\sum _{j=0}^{\infty }\frac{1}{N^j}\left[\delta (x-z_j)-\delta (x+z_j)\right] \nonumber \\
&\times &\frac{1}{\tau _0\tau ^j}\exp\left(-\frac{\Delta t}{\tau _0\tau ^j}\right) \nonumber \\
&=&\left\{
\begin{array}{cc}
\psi ^F(x,\Delta t), & \mbox{where $z_j=b_0b^j$ for CTWF}, \\
\psi ^W(x,\Delta t),~ & \mbox{where $z_j=v_0v^j\Delta t$ for CTWW}.
\end{array}
\right.
\label{rown:psixDtv}
\end{eqnarray}

The initial step of the walker requires special treatment for the stationary situations, i.e., proper averaging over initial time of the process or the performing a moving-average procedure as the time origin can be chosen arbitrarily in this situation. Applying the conditional probability techniques \cite{HauKehr0} (see also \cite{Feller}), one can define the sought stationary WTDD as follows,
\begin{eqnarray}
\chi (\Delta t)=\int_{-\infty }^{\infty }dx\chi (x,\Delta t)=\frac{\Psi (\Delta t)}{\langle \Delta t\rangle }
\label{rown:chiDt}
\end{eqnarray}
and
\begin{eqnarray}
\Psi (\Delta t)&=&\int _{-\infty }^{\infty }dx \Psi (x,\Delta t)
=\int_{\Delta t}^{\infty }d\Delta t^{\prime }\psi (\Delta t^{\prime }), \nonumber \\
\psi (\Delta t)&=&\int _{-\infty }^{\infty }dx\psi (x, \Delta t),
\label{rown:PsiDt}
\end{eqnarray}
where for sojourn probability we get,
\begin{eqnarray}
\Psi (x,\Delta t)&=&\frac{1}{2}\left(1-\frac{1}{N}\right)
\sum _{j=0}^{\infty }\frac{1}{N^j}\left[\delta (x-z_j)-\delta (x+z_j)\right]\exp\left(-\frac{\Delta t}{\tau _0\tau ^j}\right) \nonumber \\
&=&\left\{
\begin{array}{cc}
\Psi ^F(x,\Delta t), & \mbox{where $z_j=b_0b^j$ for CTWF}, \\
\Psi ^W(x,\Delta t),~ & \mbox{where $z_j=v_0v^j\Delta t$ for CTWW}.
\end{array}
\right.
\label{rown:PsixDtv}
\end{eqnarray}

From Eqs. (\ref{rown:chiDt}) -- (\ref{rown:PsixDtv}) we obtain,
\begin{eqnarray}
\chi (x,\Delta t)&=&\frac{1}{2}\left(1-\frac{\tau }{N}\right)\frac{1}{\tau _0}\sum _{j=0}^{\infty }\frac{1}{N^j}\left[\delta (x-z_j)-\delta (x+z_j)\right]\exp\left(-\frac{\Delta t}{\tau _0\tau ^j}\right) \nonumber \\
&=&\left\{
\begin{array}{cc}
\chi ^F(x,\Delta t), & \mbox{where $z_j=b_0b^j$ for CTWF}, \\
\chi ^W(x,\Delta t),~ & \mbox{where $z_j=v_0v^j\Delta t$ for CTWW}.
\end{array}
\right.
\label{rown:chixDtv}
\end{eqnarray}

Furthermore, from (\ref{rown:chixDtv}) we obtain, 
\begin{eqnarray}
\Xi (x,\Delta t)&=&\frac{1}{2}\left(1-\frac{\tau }{N}\right)\sum _{j=0}^{\infty }\left(\frac{\tau }{N}\right)^j\left[\delta (x-z_j)-\delta (x+z_j)\right]\exp\left(-\frac{\Delta t}{\tau _0\tau ^j}\right) \nonumber \\
&=&\left\{
\begin{array}{cc}
\Xi ^F(x,\Delta t), & \mbox{where $z_j=b_0b^j$ for CTWF}, \\
\Xi ^W(x,\Delta t),~ & \mbox{where $z_j=v_0v^j\Delta t$ for CTWW}.
\end{array}
\right.
\label{rown:XixDtv}
\end{eqnarray}

because the relationship between $ \chi $ and $ \Xi $, both for CTWF and CTWW, is analogous to the relationship (\ref{rown:PsiDt}) between   $\psi $ and $\Psi $.
 
 As you can see, WTDDs for CTWF and CTWW (both for non-stationary and stationary cases) differ in spatial parts. In the CTWF case, the relationship between $x$ and $\Delta t$ does not occur (see Eq. (\ref{rown:propFtermY}) for details), while in the CTWW case, there is a simple relationship (see Eq. (\ref{rown:propWtermY}) for details). This difference leads to significant differences between random walks.

\subsection{Marginal distributions}

From Eqs. (\ref{rown:psixDtv}) and (\ref{rown:PsixDtv}) -- (\ref{rown:XixDtv}) we get the list of marginal only time-dependent distributions by integrating both sides of these equations over single-step variable $x$,
\begin{eqnarray}
 \psi ^F(\Delta t)&=&\psi ^W(\Delta t) \nonumber \\
 &=&\left(1-\frac{1}{N}\right)\sum_{j=0}^{\infty }\frac{1}{N^j}\frac{1}{\tau _0\tau ^j}\exp\left(-\frac{\Delta t}{\tau _0\tau ^j}\right),
\label{rown:marginst1}
\end{eqnarray} 

\begin{eqnarray}
\Psi ^F(\Delta t)&=&\Psi ^W(\Delta t) \nonumber \\ 
&=&\left(1-\frac{1}{N}\right)\sum_{j=0}^{\infty }\frac{1}{N^j}\exp\left(-\frac{\Delta t}{\tau _0\tau ^j}\right),
\label{rown:marginst2}
\end{eqnarray}

\begin{eqnarray}
 \chi ^F(\Delta t)&=&\chi ^W(\Delta t) \nonumber \\
 &=&\left(1-\frac{\tau }{N}\right)\frac{1}{\tau _0}\sum_{j=0}^{\infty }\frac{1}{N^j}\frac{1}\exp\left(-\frac{\Delta t}{\tau _0\tau ^j}\right),
\label{rown:marginst3}
\end{eqnarray} 

\begin{eqnarray}
\Xi ^F(\Delta t)&=&\Xi ^W(\Delta t) \nonumber \\ 
&=&\left(1-\frac{1}{N}\right)\sum_{j=0}^{\infty }\left(\frac{\tau }{N}\right)^j\exp\left(-\frac{\Delta t}{\tau _0\tau ^j}\right).
\label{rown:marginst4}
\end{eqnarray} 

From Eq. (\ref{rown:psixDtv}) we get marginal space-dependent distribution by integrating both sides of above equation over single-step variable $\Delta t$,
\begin{eqnarray}
 \psi ^F(x)=\frac{1}{2}\left(1-\frac{1}{N}\right)\sum_{j=0}^{\infty }\frac{1}{N^j}\frac{1}{b_0b^j}
 \delta \left(\frac{\mid x\mid }{b_0b^j}-1\right),
\label{rown:marginsx1}
\end{eqnarray}

\begin{eqnarray}
 \psi ^W(x)=\frac{1}{2}\left(1-\frac{1}{N}\right)\sum_{j=0}^{\infty }\frac{1}{N^j}\frac{1}{b_0b^j}
 \exp \left(-\frac{\mid x\mid }{b_0b^j}\right),
\label{rown:marginsx2}
\end{eqnarray}
where $b_0=v_0\tau _0$ and $b=v\tau $. Analogously, we can easily get other distributions: $\chi ^{F,W}(x), \Psi ^{F,W}(x)$, and $\Xi^{F,W}(x)$ in  similar forms. 

\subsection{Non-factorial spatial-temporal cross-moments}\label{section:means}

We mark cross-moments and hence marginal moments for CTWF and CTWW processes both for the non-stationary and stationary cases.

\subsubsection{Non-stationary case for $n=0,1,2,\ldots $ and $m=0,1,2,\ldots $}

The spatial-temporal cross-moment takes the following form in this case,

\begin{eqnarray}
& &\langle x^{2n}\Delta t^m\rangle =\int _0^{\infty }d\Delta t\Delta t^m\int _{-\infty }^{\infty }dxx^{2n}\psi(\Delta x,\Delta t) 
\nonumber \\
&=&\left\{
\begin{array}{cc}
g(n,m)b_0^{2n}\tau _0^m\frac{1-\frac{1}{N}}{1-\frac{b^{2n}\tau ^m}{N}}, & \mbox{for $\frac{1}{\beta }<\frac{1}{2n}-\frac{m}{2n}\frac{1}{\alpha },$} \\
\infty , & \mbox{otherwise,}
\end{array}
\right.
\label{rown:x2ndtm}
\end{eqnarray}
where
\begin{eqnarray}
g(n,m)
=\left\{
\begin{array}{cc}
m!, & \mbox{for CTWF process,} \\
(2n+m)!, & \mbox{for CTWW process,}
\end{array}
\right.
\label{rown:gnm}
\end{eqnarray}

Now we can formulate the following properties.
\begin{itemize}
\item[(i)]Note that the appearance of the factor $(2n + m)!$ instead of $m!$ results directly from dependency (\ref{rown:propWtermY}).
\item[(ii)]Hence, the spatial marginal moment $\langle x^{2n}\rangle $ can be obtained by setting $m=0$, while the temporal $\langle \Delta t^m\rangle $ one by setting $n=0$.
\item[(iii)]As you can see, the cross-moment (\ref{rown:x2ndtm}) is (in general) non-factorial. This means that spatial-temporal cross-correlations reside in the CTWF and CTWW processes. The basic source of these correlations is the hierarchical nature of both processes.
\end{itemize}

\subsubsection{Stationary case for $n=0,1,2,\ldots $ and $m=0,1,2,\ldots $}\label{section:autocorr}

For the stationary case, we have different spatial-temporal cross-moments although similarly calculated,
\begin{eqnarray}
& &\langle x^{2n}\Delta t^m\rangle _{st}=\int _0^{\infty }d\Delta t\Delta t^m\int _{-\infty }^{\infty }dxx^{2n}\chi (\Delta x,\Delta t) \nonumber \\
&=&\left\{
\begin{array}{cc}
g(n,m)b_0^{2n}\tau _0^m\frac{1-\frac{\tau }{N}}{1-\frac{b^{2n}\tau ^{m+1}}{N}}, & \mbox{for $\frac{1}{\beta }<\frac{1}{2n}-\frac{m+1}{2n}\frac{1}{\alpha }$} \\
\infty , & \mbox{otherwise}
\end{array}
\right.
\label{rown:x2ndtmst}
\end{eqnarray}
where $g(n,m)$ is given by Eq. (\ref{rown:gnm}).

These cross-moments have properties analogous to those (i) -- (iii) presented above for the non-stationary case as there is a simple relation between both types of cross-moments in the form,
\begin{eqnarray}
\langle x^{2n}\Delta t^m\rangle _{st}=\frac{g(n,m)}{g(n,m+1)}\frac{\langle x^{2n}\Delta t^{m+1}\rangle }{\langle \Delta t\rangle } \nonumber \\
\ne \frac{g(n,m)}{g(n,m+1)}\langle x^{2n}\Delta t^m\rangle .
\label{rown:stnonst}
\end{eqnarray}
and 
\begin{eqnarray}
\langle x^{2n}\Delta t^m\rangle _{st}\ne \langle x^{2n}\rangle \langle \Delta t^m\rangle _{st}.
\label{rown:stnonst2}
\end{eqnarray}
Hence, we get expressions for stationary boundary moments in the form,
\begin{eqnarray}
\langle \Delta t\rangle _{st}=\frac{g(0,1)}{g(0,2)}\frac{\langle \Delta t^2\rangle }{\langle \Delta t\rangle }\ne \langle \Delta t\rangle 
\label{rown:tstnondt}
\end{eqnarray}
and
\begin{eqnarray}
\langle x^2\rangle _{st}=\frac{g(1,0)}{g(1,1)}\frac{\langle x^2\Delta t\rangle }{\langle \Delta t\rangle }\ne \langle x^2\rangle . 
\label{rown:xstnondt}
\end{eqnarray}

The cross-correlations (both for stationary and non-stationary random walks) are non-vanishing for any even function of $x$ variable. However, for odd functions of $x$ the cross-correlations vanish, because we are dealing with random walks without drifts.

\end{document}